\documentclass[a4paper, 12pt]{article}

\usepackage[a4paper,bindingoffset=0.15in,%
            left=1in,right=1in,top=1in,bottom=1in,%
            footskip=.25in]{geometry} 

\usepackage[symbol]{footmisc}
\usepackage{algorithm}
\usepackage{algpseudocode}
\usepackage{amsmath}

\usepackage{titlesec}
\usepackage{amssymb}
\usepackage{bm}
\usepackage{physics}
\usepackage{bigints}
\usepackage{bbm}
\usepackage{comment}
\usepackage{hyperref}
\newcommand{\lowerromannumeral}[1]{\romannumeral#1\relax} 
\newcommand{\upperRomannumeral}[1]{\uppercase\expandafter{\romannumeral#1}}

\usepackage{graphicx} 
\usepackage{multicol} 
\usepackage{lipsum}
\usepackage{natbib}

\usepackage{mathtools}

\newtheorem{Remark}{\bf Remark}

\usepackage[table,xcdraw,dvipsnames]{xcolor}
\usepackage{booktabs}   

\usepackage{caption}
\begin{document}

\begin{center}

{\Large \bfseries Bayesian Strategies for Repulsive Spatial \\ Point Processes}

\vspace{5 mm}

{\large Chaoyi Lu\footnote[1]{Address for correspondence: \texttt{chaoyi.lu.stat@gmail.com}}, Nial Friel} \\
School of Mathematics and Statistics, University College Dublin, Ireland\\
Insight Research Ireland Centre for Data Analytics, University College Dublin, Ireland

\vspace{5 mm}

\today 

\vspace{5mm}

\end{center}


\begin{abstract}

\noindent There is increasing interest to develop Bayesian inferential algorithms for point process models with intractable likelihoods. 
A purpose of this paper is to illustrate the utility of using simulation based strategies, including Approximate Bayesian Computation (ABC) and Markov Chain Monte Carlo (MCMC) methods for this task. 
\citet{shirota2017approximate} proposed an extended version of an ABC approach for Repulsive Spatial Point Processes (RSPP),
but their algorithm was not correctly detailed. 
In this paper, we correct their method and, based on this, 
we propose a new ABC-MCMC algorithm to which Markov property is introduced compared to a typical ABC method.
Though it is generally impractical to use, Monte Carlo approximations can be leveraged for intractable terms.
Another aspect of this paper is to explore the use of the exchange algorithm and the noisy Metropolis-Hastings
algorithm \citep{alquier2016noisy} on RSPP.
Comparisons to ABC-MCMC methods are also provided.
We find that the inferential approaches outlined above yield good performance for RSPP in both simulated and real data applications and should be considered as viable approaches for the analysis of these models.

\end{abstract}

\paragraph{Keywords and Phrases:}  Repulsive spatial point processes; Noisy Metropolis-Hastings algorithm; ABC-MCMC algorithm; Parallel computation; Strauss point process; Determinantal point process.


\section{Introduction}

A spatial point process is a random pattern of points with both the number of points and their locations being random. 
A canonical example is the Poisson point process whose number of points follows Poisson distribution with event locations being independent and identically distributed with density proportional to some intensity function. 
Such a fundamental point process model is a special case of several main classes of spatial point processes. 
These include Cox processes \citep{cox1955some}, including log Gaussian Cox processes \citep{moller1998log}, and the shot-noise Cox processes \citep{moller2003shot} for which a Neyman-Scott process \citep{cressie2015statistics} is a special case. 
Gibbs point processes (GPP) \citep{ripley1977modelling,ripley1977markov,van1995markov,moller2003statistical,illian2008statistical,gelfand2010handbook}, including Markov point processes and pairwise interaction point processes 
where a typical example is the Strauss point process (SPP). 
This is the first model which we focus on in this paper. 
The Strauss point process is an example of a doubly-intractable model where both the normalising terms of the posterior and the likelihood functions are unavailable. 
Our second model of interest is the determinantal point process (DPP) \citep{macchi1975coincidence,lavancier2015determinantal,hough2009zeros}. 
Although not strictly an intractable likelihood model, it is typically computationally expensive to evaluate. 
A determinantal point process model is usually defined over a Borel set, for example, $\mathbb{R}^d$ with $d$ being the event dimension. Simulation from such a process is required to be restricted to a specific area in practice.
The restricted version of the DPP model requires a spectral representation of the kernel and such a density function is treated as the true likelihood for the perfect sampler we focus on in our experiments.

Frequentist approaches to infer spatial point process models include those based on maximum likelihood estimation. 
However, such approaches are complicated for SPPs. 
A much quicker alternative inference method is based on the pseudo-likelihood function \citep{baddeley2000practical,jensen1991pseudolikelihood,gelfand2010handbook}, which is specified in terms of the Papangelou conditional intensity. 
The Monte Carlo Maximum Likelihood (MCML) method \citep{geyer1992constrained,geyer1994simulation,macdonald2023feasibility} is also widely used and instead replaces the intractable normalizing constant with a Monte Carlo approximation.
The MCML is asymptotically exact, and can provide more accurate inference compared to pseudo-likelihood approaches.

In this paper we focus on Bayesian approaches.
Compared to frequentist approaches which typically focus on point estimates of model parameters, Bayesian approaches instead provide the entire posterior distribution of parameters, accounting for any possibly available useful prior information included in the prior distribution.
The exchange algorithm, proposed by \citet{murray2006mcmc} for tackling doubly-intractable problems, is a natural algorithm for practitioners to consider for Gibbs-type likelihood, although it is not well explored in the context of SPPs and DPPs. 
However, in practice, it can be difficult to improve the mixing or efficiency of such an algorithm other than the computationally expensive as-long-as-possible implementations.
An auxiliary variable approach proposed by \citet{moller2006efficient} appeared around the same time as \citet{murray2006mcmc} and is also able to address Gibbs-type likelihoods.
However, it usually further requires the maximum pseudolikelihood estimates \citep{besag1974spatial} of model parameters which in turn affects the mixing performance of posterior samples.\\

An objective of this paper is to illustrate the utility of using the exchange algorithm and an extension of it, the Noisy Metropolis-Hastings (Noisy M-H) algorithm \citep{alquier2016noisy}.
The exchange algorithm uses a single draw from the likelihood function at each iteration of the Markov chain, while the Noisy M-H algorithm relies on more than one draw from the likelihood model. 
In effect, the multiple likelihood draws are used to construct a Monte Carlo estimate of the ratio of intractable likelihood normalizing constants which is then plugged into the usual Metropolis-Hastings (M-H) algorithm. 
Parallel computation can be implemented to carry out multiple draws within the Noisy M-H algorithm yielding significant improvement in computational run time and efficiency. 
The Noisy M-H algorithm is known to target the true posterior distribution when the number of likelihood draws is either one or infinity, but the convergence is not guaranteed when the number of draws is finite and greater than one.
Further, it has not yet received much attention in the context of repulsive spatial point processes, and thus we propose to explore its performance when the number of auxiliary draws is small, providing some potential efficiency.
An approximate exchange and an approximate Noisy M-H algorithm are proposed for DPPs to further facilitate the efficiency.
Comparisons to the exchange algorithm as well as approximate Bayesian computation methods are also explored. 

ABC methods are another popular class of tools to deal with the doubly-intractable problems where a closed-form likelihood function is not required during the implementations.
\citet{tavare1997inferring} first introduced ABC methods as a rejection technique in population genetics.
A generalized version was produced by \citet{pritchard1999population} where a tolerance threshold was introduced when measuring the distance between the summary statistics of observed data and those of the simulated data from the likelihood model. 
However, it instead targets an approximate posterior distribution where a smaller tolerance level leads to better accuracy. 
As such, practitioners of ABC are faced with an efficiency and accuracy trade-off when implementing such a class of methods.

\citet{marjoram2003markov} adopts an ABC-MCMC approach which can be implemented for doubly-intractable models. 
Following a similar approach, \citet{shirota2017approximate} propose an ABC approach for spatial point process models where they incorporate an ABC-like rejection sampling step as the proposal step of the MCMC scheme.
A semi-automatic approach proposed by \citet{fearnhead2012constructing} is further adopted by \citet{shirota2017approximate} in order to find an optimal choice of the summary statistic of the observed data for model parameters.
However, we show in Section~\ref{ABC section} that the proposed ABC-MCMC algorithm in \citet{shirota2017approximate} leads to a stationary distribution which is different from the target posterior distribution with an intractable multiplicative term that depends on model parameters, a fact apparently overlooked by the authors.
We propose to correct their method leading to a new ABC-MCMC algorithm with intractable multiplicative terms included inside the acceptance ratio.
In order to explore this corrected version which is impossible to implement in general, we estimate those intractable terms by leveraging Monte Carlo approximations that can be easily parallelized.
An approximate parallel computation is also proposed to implement the \texttt{repeat} loop embedded inside the algorithm.
The resulting corrected \citet{shirota2017approximate} ABC-MCMC algorithm is compared to the \citet{fearnhead2012constructing} ABC-MCMC algorithm as well as the exchange and Noisy M-H algorithms outlined above.
All the four algorithms mentioned here rely on simulation from intractable likelihood model. \\

It is worthwhile to notice that, in the literature, \citet{park2018bayesian} also compares the Noisy M-H algorithm and the exchange algorithm as well as several other auxiliary-draw and likelihood-approximation MCMC methods on various types of models.
However, due to the fact that perfect sampling of the spatial point process model they worked on is not available, a pseudo-marginal MCMC method, an approximate exchange algorithm along with a noisy version of it are instead compared.
Further, the comparisons between MCMC methods and ABC-MCMC algorithms are not well explored for RSPP in the literature.
Thus our contribution also aims to fulfil these gaps by focusing on two basic RSPP models, namely, SPPs and DPPs, for which perfect simulation is possible.

The structure of this paper is as follows. 
Section~\ref{SectionRSPP} provides a basic introduction of SPPs and DPPs.
The MCMC methods including the M-H algorithm, the exchange algorithm, the Noisy M-H algorithm as well as our proposed approximate exchange algorithm and approximate Noisy M-H algorithm for DPPs are illustrated in Section~\ref{MCMC}.
In Section~\ref{ABC section}, an overview of the ABC-MCMC algorithm proposed by \citet{shirota2017approximate} is included, and we detail the incorrect stationary distribution that their method target.
Based on the correction of the \citet{shirota2017approximate} ABC-MCMC algorithm, a new ABC-MCMC algorithm is proposed therein.
Simulation studies and real data applications are demonstrated, respectively, in Sections \ref{SS} and \ref{RDA}. 
Conclusions and discussions follow in Section \ref{Conclusion}.


\section{Repulsive Spatial Point Processes}
\label{SectionRSPP}

In this section, we introduce two basic repulsive spatial point process models we work on for algorithm explorations in this paper. 
The first one is a Strauss Point Process which is a specific case of the Gibbs Point Processes, whereas the second one is a Determinantal Point Process with a Gaussian kernel. 


\subsection{Strauss Point Processes}
\label{SectionSPP}

A finite point process $X$ with realization $\bm{x}:= \{x_1,x_2,\dots,x_N\}$ from a finite point configuration space $\bm{N}_f:=\{\bm{x}\subset S:N<\infty\}$ on a bounded spatial domain $S \subset \mathbb{R}^d$ is said to be a Gibbs point process (GPP) if it admits a density $f(\bm{x})\propto \text{exp}\left(\sum_{\emptyset\neq\bm{y}\subseteq\bm{x}}Q(\bm{y})\right)$ with respect to a homogeneous Poisson point process with unit intensity. 
Setting $\text{exp}(-\infty)=0$, the potential $Q(\bm{x})\in[-\infty,\infty)$ is defined for all non-empty finite point patterns $\bm{x} \subset S$. Conditional on $N = n$, the joint probability density of $\bm{x}$ is of the form
\begin{equation*}
\label{GPPDensity}
\begin{split}
p_n(x_1,x_2,\dots,x_n|N=n)
&\propto \frac{\text{exp}(-\mu(S))}{n!}\mu(S)^n f(\{x_1,\dots,x_n\}),\\ 
\end{split}
\end{equation*}
where $\mu(\cdot)$
is the corresponding intensity measure on $S$, and the normalising term forms the distribution of $N$, that is,
\begin{equation*}
\label{GPPDensityNomalisation}
\begin{split}
\text{P}(N=n)
&= \frac{\text{exp}(-\mu(S))}{n!} \int_S\dots\int_Sf(\{x_1,\dots,x_n\})\text{d}\mu(x_1)\dots \text{d}\mu(x_n).\\ 
\end{split}
\end{equation*}
In most cases, $f$ is specified up to a proportionality constant and we denote $f\propto h$ where $h:\textbf{N}_f\rightarrow [0,\infty)$ is a known function. 

A homogeneous pairwise interaction point process is a GPP with density
\begin{equation*}
\label{GPPDensity}
\begin{split}
f(\bm{x})
&\propto h(\bm{x}) := \beta^{n(\bm{x})}\prod_{\{u,v\}\subseteq \bm{x}}\psi (\{u, v\}),\\ 
\end{split}
\end{equation*}
where $\beta > 0$ is a constant and $n(\bm{x})$ is the number of events in $\bm{x}$. Moreover, $\psi : S \times S \rightarrow [0,\infty)$ is an interaction function defined as $\psi(\{u, v\}):=\psi_0(||u-v||)$ where $\psi_0: (0, \infty)\rightarrow [0,\infty)$ is invariant under reflections, rotations and translations. The Strauss point process (SPP) is the simplest non-trivial homogeneous pairwise interaction process with
\begin{equation*}
\begin{split}
\psi_0(||u-v||)
&= \gamma^{\mathbbm{1}(||u-v|| \leq R)}.\\ 
\end{split}
\end{equation*}
The density of the Strauss point process is of the form 
\begin{equation}
\label{SPPDensity}
\begin{split}
f(\bm{x}) 
&\propto h(\bm{x}) := \beta^{n(\bm{x})}\gamma^{s_R(\bm{x})},\\ 
\end{split}
\end{equation}
where $\beta>0, 0\leq \gamma \leq 1$ and 
\begin{equation*}
\begin{split}
s_R(\bm{x})
&:= \sum_{ \{ u, v \} \subseteq \bm{x} } \mathbbm{1}(||u - v|| \leq R), \ \ u, v \in S,\\ 
\end{split}
\end{equation*}
which is the number of $R$-close pairs of points in $\bm{x}$. Further, $\mathbbm{1}(\cdot)$ is the indicator function, $\beta$ is known as the rate and $\gamma$ is the interaction parameter where smaller values of $\gamma$ implies stronger repulsion. 
We refer to, for example, \citet{ripley1977modelling,ripley1977markov,van1995markov,moller2003statistical,illian2008statistical,gelfand2010handbook} for more details of the model.
Perfect simulation of a SPP can be accomplished by applying the ``dominated coupling from the past'' algorithm \citep{kendall2000perfect} which can be implemented in the R package \texttt{spatstat} \citep{baddeley2005spatstat}.


\subsection{Determinantal Point Processes}
\label{DPPG}

A simple locally finite spatial point process $X$ on $\mathbb{R}^2$ is called a determinantal point process with kernel $C$ \citep{lavancier2015determinantal} if it has a product density function
\begin{equation}
\label{DPPDensity}
\begin{split}
\rho^{(n)}(x_1,\dots,x_n)
&= \text{det}[C](x_1,\dots,x_n),\\ 
\end{split}
\end{equation}
where $(x_1,\dots,x_n) \in (\mathbb{R}^2)^n$ for $n = 1,2, \dots$, and $[C](x_1,\dots,x_n)$ is the $n \times n$ matrix with $(i,j)$th entry being $C(x_i,x_j)$. 
Here $C$ is a covariance kernel defined on $\mathbb{R}^2\times\mathbb{R}^2$ and $\text{det}(\cdot)$ denotes determinant of the matrix. 
We write $X \sim$ DPP$_{\mathbb{R}^2}(C)$ and, for any compact subset $S \subseteq \mathbb{R}^2$, we denote by $\text{DPP}_S(C)$, the distribution of the DPP on $S$ with kernel given by the restriction of $C$ to $S \times S$. 
Note that $\text{DPP}_S(C)$ is equivalent to the distribution of $X_S=X\cap S$. 

The intensity function is the first order density function: $\rho^{(1)}(x) = C(x,x)$ for $x \in \mathbb{R}^2$ and where the pairwise correlation function of $X$ is
$$g(x,y) = \frac{\rho^{(2)}(x,y)}{\rho^{(1)}(x)\rho^{(1)}(y)} = 1 - \frac{C(x,y)C(y,x)}{C(x,x)C(y,y)},$$
if $C(x,x) > 0$ and $C(y,y) > 0$. Otherwise it is equal to zero. 
As $C$ is a real covariance kernel in our setting, the repulsiveness of the DPP is reflected by $g \leq 1$ and  
\begin{equation*}
\begin{split}
\rho^{(n)}(x_1,\dots,x_n)
&\leq \rho^{(1)}(x_1)\cdots \rho^{(1)}(x_n),\\ 
\end{split}
\end{equation*}
for any $n = 2,3,\dots$. The inequality follows from the fact that the determinant of a covariance matrix never exceeds the product of its diagonal elements. We refer to \citet{hough2006determinantal} and \citet{lavancier2015determinantal} for more details and properties of DPPs.\\

Since simulation from DPPs over $\mathbb{R}^2$ is practically impossible, \citet{lavancier2015determinantal} instead focused on the restricted $X_{S}$ within a bounded region $S \subseteq \mathbb{R}^2$.
The kernel $C$ restricted to $S \times S$ has a spectral representation: $C(x,y) = \sum^\infty_{k=1}\lambda_k\phi_k(x)\overline{\phi_k(y)}$ where $(x,y) \in S \times S$, and $\overline{(\cdot)}$ denotes the complex conjugate.
Here, $\{\phi_k\}^\infty_{k=1}$ are eigenfunctions and $\{\lambda_k\}^\infty_{k=1}$ are the set of eigenvalues which are required to be $\leq 1$ in order to ensure the existence of DPPs. 
For stationary continuous DPPs, the density of $X_{S}$, which has realizations $(x_1,\dots,x_n)\in S^n$, is of the form
\begin{equation}
\label{DPPDensityS}
\begin{split}
f(x_1,\dots,x_n)
&= \text{exp}(|S|-D)\text{det}[\tilde{C}](x_1,\dots,x_n),\\ 
\end{split}
\end{equation}
for $n = 0,1,2,\dots$, where $D=-\text{logP}(n=0)=-\sum_{k=1}^\infty \text{log}(1-\lambda_k)$ and $\tilde{C}(x,y)=\sum^\infty_{k=1}\frac{\lambda_k}{1-\lambda_k}\phi_k(x)\overline{\phi_k(y)}$ with the assumption that $\text{det}[\tilde{C}](x_1,\dots,x_n)=1$ if $n =0$. 

However, it is not always the case that the spectral representation of the covariance kernel is explicitly known. Thus \citet{lavancier2015determinantal} proposed to apply a Fourier series approximation of the kernel and instead consider the approximate DPP: $\hat{X}_S\sim \text{DPP}_S(\hat{C})$ where
\begin{equation}
\label{DPPApproxKernelS}
\begin{split}
\hat{C}(x,y)
&= \sum_{k\in \mathbb{Z}^2}\varphi(k)\text{exp}(2\pi ik\cdot(x-y)),\hspace{1em} x,y \in S.\\ 
\end{split}
\end{equation}
Defining $C(x,y)=C_0(x-y)$, then $\varphi(k)$, which is called spectral density, is the Fourier transform of $C_0$. 
The corresponding approximate density is of the form
\begin{equation}
\label{DPPApproxDensityS}
\begin{split}
\hat{f}(x_1,\dots,x_n)
&=  \text{exp}(|S|-\hat{D})\text{det}[\tilde{\hat{C}}](x_1,\dots,x_n),\\ 
\end{split}
\end{equation}
where $\tilde{\hat{C}}(x,y)=\sum_{k\in \mathbb{Z}^2}\tilde{\varphi}(k)\text{exp}(2\pi ik\cdot(x-y))$, $\hat{D}=\sum_{k\in \mathbb{Z}^2}\text{log}(1+\tilde{\varphi}(k))$ and $\tilde{\varphi}(k) = \varphi(k)/(1-\varphi(k))$. 
Nevertheless, summation over the whole $\mathbb{Z}^2$ here is impossible in practice, so a truncation $M$ is applied by \citet{lavancier2015determinantal}, so that $\sum_{k\in \mathbb{Z}_M^2}\varphi(k)>0.99n/|S|$, where $\mathbb{Z}_M = \{-M,-M+1,\dots,M-1,M\}$. 
Thus the simulation algorithm introduced therein can be treated as a perfect sampler from the likelihood function $\hat{f}$ shown in Eq.~\eqref{DPPApproxDensityS} with the truncation approximation applied, and is available in the \texttt{spatstat} R package.
We refer to \citet{lavancier2015determinantal} for more details.\\

In this paper, apart from the SPPs, we also work on the approximate restricted DPP$_S(\hat{C})$ with the likelihood function $\hat{f}$ for algorithm explorations.
The covariance kernel is proposed to be the Gaussian kernel which is stationary and isotropic.
The Gaussian kernel and its corresponding spectral density are, respectively, written as
\begin{equation}
\label{DPPGaussianKernel}
\begin{split}
C(x,y) =  \tau\text{exp}(-||x-y||^2/\sigma^2),\;\;\;\varphi(k)=\tau(\sqrt{\pi}\sigma)^2\text{exp}(-||\pi\sigma k||^2),
\end{split}
\end{equation}
where $\tau$ is the intensity parameter corresponding to the expected number of points per unit of area, and $\sigma$ is a scale parameter of the kernel.
The existence of this Determinantal Point Process with a Gaussian kernel (DPPG) is guaranteed by $0\leq \sigma \leq \sigma_{max}=1/\sqrt{\pi \tau}$ or, equivalently, $0\leq \tau \leq \tau_{max}=1/(\pi \sigma^2)$.


\section{Markov Chain Monte Carlo}
\label{MCMC}

We consider a posterior distribution $\pi(\theta|y) = \mathcal{L}(y|\theta) \pi(\theta)/z(y)$ with a likelihood function of the form $\mathcal{L}(y|\theta)= q(y|\theta)/z(\theta)$, where $y\in \mathcal{Y}$ and $\theta\in\Theta$ are the observed data and model parameter(s), respectively. 
The terms $z(y)$ and $z(\theta)$ are the normalizing constant of the posterior and the likelihood, respectively.
We denote $q(\cdot|\theta)$ as the unnormalized likelihood function, whereas the prior distribution for $\theta$ is given by $\pi(\theta)$.
Under these settings, the Metropolis-Hastings algorithm outlined in Algorithm~\ref{MH}, which is a basic Markov Chain Monte Carlo (MCMC) approach to exactly sample from the posterior distribution, does not apply when the normalizing term $z(\theta)$ is intractable.

\begin{algorithm}[h!]
\caption{Metropolis-Hastings algorithm}
\label{MH}
\begin{algorithmic} 
\State First initialize $\theta_0$.
\For {$t=1$ to $T$}
\State 1. Propose $\theta'$ from proposal distribution $p(\cdot|\theta^{(t-1)})$.
\State 2. Set $\theta^{(t)}=\theta'$ with probability:
\begin{align}
\label{eqn:mh}
 \alpha_{\scalebox{.85}{$\scriptscriptstyle MH$}} 
&=\min\left(1,\frac{\pi(\theta'|y)}{\pi(\theta^{(t-1)}|y)}\frac{p(\theta^{(t-1)}|\theta')}{p(\theta'|\theta^{(t-1)})}\right) \nonumber\\
 &=\min\left(1,\frac{q(y|\theta')z(\theta^{(t-1)})\pi(\theta')}{q(y|\theta^{(t-1)})z(\theta')\pi(\theta^{(t-1)})}\frac{p(\theta^{(t-1)}|\theta')}{p(\theta'|\theta^{(t-1)})}\right). 
 \end{align}
\hspace{2.4em} Otherwise, set $\theta^{(t)}=\theta^{(t-1)}$.
\EndFor
\end{algorithmic}
\end{algorithm}


\subsection{The exchange algorithm and the noisy Metropolis-Hastings algorithm}
\label{Section_ex_noisyMH}

In order to deal with doubly-intractable problems, the well-known exchange algorithm was proposed by \citet{murray2006mcmc} and is shown as Algorithm~\ref{Ex}.
The requirement of perfect sampling from the likelihood is satisfied for both SPP and DPPG as we discussed in Section~\ref{SectionRSPP}.
This algorithm instead samples from an augmented distribution $\pi(\theta,\theta', x'|y) \propto \mathcal{L}(y|\theta)\pi(\theta)p(\theta'|\theta)\mathcal{L}(x'|\theta')$, where the auxiliary function $\mathcal{L}(x'|\theta')=q(x'|\theta')/z(\theta')$ has the same form as the likelihood $\mathcal{L}(y|\theta)=q(y|\theta)/z(\theta)$.
This leads to a cancellation of the normalizing constants $z(\theta^{(t-1)})$, $z(\theta')$ in the acceptance ratio $\alpha_{\scalebox{.85}{$\scriptscriptstyle Ex$}}$ of Algorithm~\ref{Ex}.

\begin{algorithm}[H]
\caption{Exchange algorithm}
\label{Ex}
\begin{algorithmic} 
\State First initialize $\theta^{(0)}$.
\For {$t=1$ to $T$}
\State 1. Propose $\theta' \sim p(\cdot|\theta^{(t-1)})$. 
\State 2. Simulate $x' \sim \mathcal{L}(\cdot|\theta')$.
\State 3. Set $\theta^{(t)} = \theta'$ with probability:
\begin{align} 
\label{eqn:exchange}
 \alpha_{\scalebox{.85}{$\scriptscriptstyle Ex$}} 
 &= \min\left(1, \frac{q(y|\theta')}{q(y|\theta^{(t-1)})} \frac{\pi(\theta')p(\theta^{(t-1)}|\theta')}{\pi(\theta^{(t-1)})p(\theta'|\theta^{(t-1)})} \frac{q(x'|\theta^{(t-1)})}{q(x'|\theta')} \right). 
\end{align}
\hspace{2.4em} Otherwise, set $\theta^{(t)} = \theta^{(t-1)}$.
\EndFor
\end{algorithmic}
\end{algorithm}
Notice that the auxiliary unnormalized likelihood ratio $q(x'|\theta^{(t-1)})/q(x'|\theta')$ in Eq.~\eqref{eqn:exchange} of Algorithm~\ref{Ex} is actually an unbiased importance sampling estimator of the normalizing constant ratio $z(\theta^{(t-1)})/z(\theta')$ in  Eq.~\eqref{eqn:mh} of Algorithm~\ref{MH}, that is,
\[\frac{z(\theta^{(t-1)})}{z(\theta')} = \mathbb{E}_{x' \sim \mathcal{L}(\cdot|\theta')}\left(\frac{q(x'|\theta^{(t-1)})}{q(x'|\theta')}\right) \approx \frac{1}{K}\sum_{k=1}^K \frac{q(x_k'|\theta^{(t-1)})}{q(x_k'|\theta')},\]
where an improved unbiased estimator of the normalizing term ratio can be obtained by applying a simple Monte Carlo estimator of the expectation.
Here, an set of \textit{i.i.d.} $(x_1', x_2', \dots, x_K')$ is  sampled from $ \mathcal{L}(\cdot|\theta')$ for the approximation.
This estimator can be plugged into Eq.~\eqref{eqn:mh} yielding the noisy Metropolis-Hastings algorithm \citep{alquier2016noisy} as shown in Algorithm~\ref{NMH}. 
Parallel computation can be applied to sample $K$ auxiliary chains via the \texttt{doParallel} R package for more efficient implementations.

\begin{algorithm}[H]
\caption{Noisy Metropolis-Hastings algorithm}
\label{NMH}
\begin{algorithmic} 
\State First initialise $\theta^{(0)}$.
\State \textbf{for} {$t=1$ to $T$} \textbf{do}
\begin{enumerate}
\item Propose $\theta' \sim p(\cdot|\theta^{(t-1)})$.
\item \textbf{for} $k = 1, \dots, K$ \textbf{do} Generate $x_k' \sim \mathcal{L}(\cdot|\theta')$. \textbf{end for}
\item Set $\theta^{(t)}=\theta'$ with probability:
$$\alpha_{\scalebox{.85}{$\scriptscriptstyle NMH$}} = \text{min}\left(1, \frac{q(y|\theta')}{q(y|\theta^{(t-1)})} \frac{\pi(\theta')p(\theta^{(t-1)}|\theta')}{\pi(\theta^{(t-1)})p(\theta'|\theta^{(t-1)})}  \cdot  \frac{1}{K} \sum_{k=1}^K \frac{q(x_k'|\theta^{(t-1)})}{q(x_k'|\theta')}\right).$$
Otherwise, set $\theta^{(t)} = \theta^{(t-1)}$.
\end{enumerate}
\textbf{end for}
\end{algorithmic}
\end{algorithm}

\begin{Remark}
\normalfont Note that for $K = 1$, the Noisy M-H algorithm is equivalent to the exchange algorithm and, as $K \rightarrow \infty$, the algorithm becomes the standard Metropolis-Hastings algorithm. 
This indicates that both $K = 1$ and $K \rightarrow \infty$ leave the target distribution invariant, but this is not guaranteed for $1 < K < \infty$. 
\citet{alquier2016noisy} proved that a Markov chain resulting from the Noisy M-H algorithm will converge to the target posterior density as $K \rightarrow \infty$ under certain assumptions:
\begin{algorithmic}
\Statex (\textbf{A1}) the Markov chain yielded by Algorithm~\ref{MH} is uniformly ergodic.
\Statex (\textbf{A2}) there exists a constant $c_{\pi}$ such that $1/c_{\pi} \leq \pi(\theta) \leq c_{\pi}.$
\Statex (\textbf{A3}) there exists a constant $c_{p}$ such that $1/c_{p} \leq p(\theta'|\theta) \leq c_{p}.$
\Statex (\textbf{A4}) for any $\theta^{(t-1)}$ and $\theta'$, $\text{var}_{y'\sim \mathcal{L}(\cdot|\theta')}\left(q(y'|\theta^{(t-1)})/q(y'|\theta')\right) < +\infty$.
\end{algorithmic}
\citet{alquier2016noisy} also provide results for the case where the assumption of uniform ergodicity is replaced by the less restrictive assumption of geometric ergodicity.
We refer to \citet{mitrophanov2005sensitivity} and \citet{alquier2016noisy} for more details.
In what follows we assume that assumption (\textbf{A1}) holds, although this may be difficult to prove in practice. 
\label{Remark_4assumptions}
\end{Remark}
\begin{Remark}
\normalfont Both (\textbf{A2}) and (\textbf{A3}) are satisfied when one uses bounded proposal and prior distributions. 
Further, \citet{alquier2016noisy} showed that (\textbf{A4}) is satisfied for Gibbs random fields for which the SPP is a specific case: considering $h(\bm{x}) = \beta^{n(\bm{x})}\gamma^{s_R(\bm{x})}=\text{exp}(n(\bm{x})\text{log}(\beta)+s_R(\bm{x})\text{log}(\gamma))$. 
The assumption (\textbf{A4}) also holds for a DPP$_S(\hat{C})$ by noticing that $\text{exp}(|S|-\hat{D})\leq \hat{f} \leq \text{exp}(|S|-\hat{D})(\sum_{k\in \mathbb{Z}^2}\tilde{\varphi}(k))^n$ for any $n=0,1,2,\dots$, where $\sum_{k\in \mathbb{Z}_M^2}\varphi(k)$ tends to $\tau$ from below as truncation $M\rightarrow \infty$. 
Thus, assuming that all four assumptions in \textbf{Remark}~\ref{Remark_4assumptions} hold, then Theorem $3.1$ of \citet{alquier2016noisy} provides a theoretical guarantee that the Markov chain resulting from the Noisy M-H algorithm will converge to the target posterior density as $K \rightarrow \infty$. 
However, in practice one applies the Noisy M-H algorithm with a finite number, $K$, of auxiliary chains. 
Especially when $K$ is small, the lower computational burden is appealing.
The objective in this case is to explore the accuracy performance of the small-$K$ Noisy M-H algorithm as well as the potential improvement in both mixing and efficiency performance compared to the exchange algorithm.
By treating the exchange algorithm as one of our benchmarks, we investigate this possible accuracy-efficiency trade-off in this paper.\\
\end{Remark}


\subsection{An approximate noisy Metropolis-Hastings algorithm for DPPG}
\label{Section_approx_noisyMH_DPPG}

Recall that the $\hat{X}_S\sim\text{DPP}_S(\hat{C})$ with a Gaussian kernel shown as Eq.\eqref{DPPApproxDensityS} and Eq.\eqref{DPPGaussianKernel} is one of the two point process models we focus on for algorithm performance investigations.
\citet{lavancier2015determinantal} shows that realizations simulated from $\hat{f}$ in Eq.\eqref{DPPApproxDensityS} provides the empirical means of $L(r)-r$ being close to the corresponding theoretical $L(r)-r$ function, where the $L$-function is the variance stabilizing transformation of the $K$-function which is defined as $K(r):= \pi r^2-(1-\text{exp}(-2r^2/\sigma^2))\pi \sigma^2/2$ for the Gaussian model. 
This indicates that the simulations from $\hat{f}$ are appropriate approximations of the DPP$_S(C)$ with a Gaussian kernel.

In contrast to the SPP, the normalizing term of $\hat{f}$ is analytically available when the truncation is applied, so it may seem unnecessary to implement the exchange algorithm or the Noisy M-H algorithm, since the Metropolis-Hastings algorithm is available. 
However, the normalizing term as well as the covariance kernel $\hat{C}$ between each pair of events in the likelihood function can be computationally expensive when $M$ is large in the truncation approximation setting. 
Thus, in order to significantly reduce the computational burden, we propose an approximate Noisy M-H algorithm by leveraging Eq.~\eqref{DPPDensity} as an approximate density of $\hat{X}_S$.

More specifically, we first denote
\begin{equation}
\label{DPPqapp}
\begin{split}
\hat{q}(x_1,\dots,x_n)
&=  \text{det}[\tilde{\hat{C}}](x_1,\dots,x_n),\\ 
\end{split}
\end{equation}
as the unnormalized likelihood function of a $\text{DPP}_S(\hat{C})$ with a Gaussian kernel, and the standard Noisy M-H algorithm for this DPPG is shown as Algorithm~\ref{NMHforDPP}, which can be used to infer the parameters $\bm{\theta} = (\tau,\sigma)$.
The DPP$_S(\hat{C}')$ in Step~2 corresponds to the model with the proposed model parameters $\bm{\theta}' = (\tau',\sigma')$.
The exchange algorithm is the $K=1$ specific case of such an algorithm. 
The target posterior distribution is thus of the form
\begin{equation*}
\label{DPPappPosterior}
\begin{split}
\pi(\tau,\sigma|\bm{y})
&\propto  \hat{q}(\bm{y}|\tau,\sigma)\pi(\tau,\sigma).
\end{split}
\end{equation*}
Next, we define $\rho(\bm{x}):=\rho^{(n)}(x_1,\dots,x_n)$, for $n=1,2,\dots$, and then, for any $\bm{x}\in X_S$, we consider the truncated density of the form
\begin{equation}
\label{DPPProportionalDensity}
\begin{split}
f(\bm{x}|\bm{x}\in X_S)
&=  \mathbbm{1}(\bm{x}\in X_S)\cdot\frac{\rho(\bm{x})}{\text{P}(\bm{x}\in X_S)}\propto \rho(\bm{x}), 
\end{split}
\end{equation}
to substitute all the $\hat{q}(\cdot)$ in the $\alpha_{\scalebox{.85}{$\scriptscriptstyle DPPG$}}$ of Algorithm~\ref{NMHforDPP}.
This leads to the approximate Noisy M-H algorithm for DPPGs illustrated as Algorithm~\ref{ApproximateNMHforDPP}.
Here, $\mathbbm{1}(\bm{x}\in X_S)$ is the indicator function returning $1$ if $\bm{x}\in X_S$ and $0$ otherwise.
The term $\text{P}(\bm{x}\in X_S)$ in Eq.\eqref{DPPProportionalDensity} is the normalizing constant over all the possible events of $X_S$. 
The approximate exchange algorithm is the $K=1$ case of Algorithm~\ref{ApproximateNMHforDPP}.
\begin{algorithm}[h!]
\caption{Noisy Metropolis-Hastings algorithm for DPP$_S(\hat{C})$}
\label{NMHforDPP}
\begin{algorithmic} 
\State First initialise $\bm{\theta}^{(0)}$.
\State \textbf{for} {$t=1$ to $T$} \textbf{do}
\begin{enumerate}
\item Propose $\bm{\theta}' \sim p(\cdot|\bm{\theta}^{(t-1)})$.
\item \textbf{for} $k = 1, \dots, K$ \textbf{do} 
Generate $\bm{x}_k' \sim$ DPP$_S(\hat{C}')$.
\textbf{end for}
\item Set $\bm{\theta}^{(t)}=\bm{\theta}'$ with probability:
$$\alpha_{\scalebox{.85}{$\scriptscriptstyle DPPG$}} = \text{min}\left(1, \frac{\hat{q}(\bm{y}|\bm{\theta}')}{\hat{q}(\bm{y}|\bm{\theta}^{(t-1)})}\frac{\pi(\bm{\theta}')p(\bm{\theta}^{(t-1)}|\bm{\theta}')}{\pi(\bm{\theta}^{(t-1)})p(\bm{\theta}'|\bm{\theta}^{(t-1)})}  \cdot  \frac{1}{K} \sum_{k=1}^K \frac{\hat{q}(\bm{x}_k'|\bm{\theta}^{(t-1)})}{\hat{q}(\bm{x}_k'|\bm{\theta}')}\right).$$
Otherwise, set $\bm{\theta}^{(t)} = \bm{\theta}^{(t-1)}$.
\end{enumerate}
\textbf{end for}
\end{algorithmic}
\end{algorithm}

\begin{algorithm}[h!]
\caption{Approximate noisy Metropolis-Hastings algorithm for DPP$_S(\hat{C})$}
\label{ApproximateNMHforDPP}
\begin{algorithmic} 
\State First initialise $\bm{\theta}^{(0)}$.
\State \textbf{for} {$t=1$ to $T$} \textbf{do}
\begin{enumerate}
\item Propose $\bm{\theta}' \sim p(\cdot|\bm{\theta}^{(t-1)})$.
\item \textbf{for} $k = 1, \dots, K$ \textbf{do} 
Generate $\bm{x}_k' \sim$ DPP$_S(\hat{C}')$. 
\textbf{end for}
\item Set $\bm{\theta}^{(t)}=\bm{\theta}'$ with the approximate acceptance ratio:
 \begin{equation}
\label{DPPNMHApproximateAcceptanceRatio}
\begin{split}
\widetilde{\alpha}_{\scalebox{.85}{$\scriptscriptstyle DPPG$}}
&=  \text{min}\left(1, \frac{\rho(\bm{y}|\bm{\theta}')}{\rho(\bm{y}|\bm{\theta}^{(t-1)})} \frac{\pi(\bm{\theta}')p(\bm{\theta}^{(t-1)}|\bm{\theta}')}{\pi(\bm{\theta}^{(t-1)})p(\bm{\theta}'|\bm{\theta}^{(t-1)})} \cdot  \frac{1}{K} \sum_{k=1}^K \frac{\rho(\bm{x}_k'|\bm{\theta}^{(t-1)})}{\rho(\bm{x}_k'|\bm{\theta}')}\right).\\ 
\end{split}
\end{equation}
Otherwise, set $\bm{\theta}^{(t)} = \bm{\theta}^{(t-1)}$.
\end{enumerate}
\textbf{end for}
\end{algorithmic}
\end{algorithm}
Note that the approximate Noisy M-H algorithm does not require knowledge of the specific form of the $\text{P}(\bm{x}\in X_S)$ in Eq.\eqref{DPPProportionalDensity} due to the cancellation of the normalizing constants in the $\widetilde{\alpha}_{\scalebox{.85}{$\scriptscriptstyle DPPG$}}$. 
Further, instead of the computationally expensive approximate kernel shown in Eq.\eqref{DPPApproxKernelS},
directly computing the Gaussian kernel from Eq.\eqref{DPPGaussianKernel}
significantly reduces the computation.
In the meanwhile, this approximation also does not lose much accuracy as $\hat{C}$ is an appropriate approximation of $C$ following the experiments and discussion in \citet{lavancier2015determinantal}. 
The performance comparisons to benchmark algorithms as well as other candidate algorithms 
are illustrated in Section~\ref{SS2DPPG}. 
Similar MCMC approximation schemes can also be found, for example, in \citet{murray2012bayesian}, but what is different to our paper is that, instead of approximating the unnormalized likelihood, they proposed a variety of approximations for the likelihood normalizing constants or for the corresponding constant ratios. 
The motivations behind these approximations are similar to the interpretation of the unbiased importance sampling estimator of the normalizing constant ratio we discussed in Section \ref{Section_ex_noisyMH} for the exchange algorithm.


\section{ABC Algorithms for Repulsive Spatial Point Processes}
\label{ABC section}

The approximate Bayesian computation approaches are also able to tackle doubly-intractable problems, but they instead target an approximate posterior distribution.
However, such an approximate distribution can be adjusted by a tuning parameter to be closer to the true posterior distribution. 
An important ingredient of ABC algorithms is the requirement to obtain a realization $x' \sim \mathcal{L}(\cdot|\theta')$, for model parameter(s) $\theta'$. Additionally, one requires a summary statistic $\bm{T}(\cdot)$ and a distance function $\Psi(\bm{T}(x'), \bm{T}(y))$, which describes the discrepancy, in some sense, between the observed data $y$ and the realization $x'$.

One of the first ABC algorithms was proposed by \citet{pritchard1999population} and is shown as Algorithm \ref{ABC}. 
In practice, the acceptance threshold $\epsilon$ will not necessarily be set to zero. 
Therefore this algorithm will not target the true posterior distribution and will instead sample from an approximate posterior distribution $\pi_{\epsilon}(\theta|y)$, obtained by marginalising over $x$, the joint distribution $\pi_{\epsilon}(\theta,x|y) \propto \mathcal{L}(x|\theta)\pi(\theta)\mathbbm{1}(\Psi(\bm{T}(x),\bm{T}(y))\leq\epsilon)$. 
The need for a small $\epsilon$ to ensure the accuracy of the resulting ABC target distribution to the posterior distribution should be balanced with the requirement that the algorithm mixes sufficiently.
Implementing ABC methods also require perfect simulation from the intractable likelihood function. 
\begin{algorithm}[H]
\caption{ABC algorithm}
\label{ABC}
\begin{algorithmic} 
\For {$t=1$ to $T$}
\State 1. Generate $\theta' \sim \pi(\theta)$ and $x' \sim \mathcal{L}(\cdot|\theta')$. Repeat this step until $\Psi(\bm{T}(x'),\bm{T}(y))\leq\epsilon$.
\State 2. Set $\theta^{(t)}=\theta'$. 
\EndFor
\end{algorithmic}
\end{algorithm}

\citet{shirota2017approximate} proposed an ABC-MCMC-like algorithm for repulsive spatial point processes.
This method is based on the likelihood-free MCMC idea explored in the ABC-MCMC algorithm proposed by \citet{marjoram2003markov} as well as a semi-automatic approach proposed by \citet{fearnhead2012constructing}. 
Such an approach argues that the optimal choice of the summary statistic $\bm{T}(\bm{y})$ in ABC methods is $\mathbb{E}(\bm{\theta}|\bm{y})$ and a linear regression scheme is considered to construct this summary statistic. 
Taking a SPP model as an example here, $\{(\beta_l,\gamma_l)\}^L_{l=1}$ are generated from prior distributions through a pilot run with each of the corresponding realizations $\{\bm{x}_l\}^L_{l=1}$ generated from the SPP likelihood function $\mathcal{L}(\cdot|\beta_l,\gamma_l)$ (\ref{SPPDensity}). 
The radius $R = \hat{R}$ is estimated by profile pseudo-likelihood method \citep{shirota2017approximate,baddeley2000practical}. 
Note that taking a log transform of the parameters can facilitate the regression, and thus the pilot draws $\{\bm{\theta}_l\}^L_{l=1}$ in the regression are stored as $\{(\text{log}(\beta_l),\text{log}(\gamma_l))\}^L_{l=1}$. 
The linear regression is implemented for $\mathbb{E}(\bm{\theta}_l|\bm{y}) = \bm{a} + \bm{b}\bm{\eta}(\bm{x}_l, \bm{y})$, where $\bm{\eta}(\bm{x}_l, \bm{y})$ is a vector of summary statistics
$$\bm{\eta}(\bm{x},\bm{y}) = (\eta_1(\bm{x},\bm{y}), \eta_2(\bm{x},\bm{y})),$$ 
with
$$\eta_1(\bm{x},\bm{y}) = \text{log}(n(\bm{x})) - \text{log}(n(\bm{y})), \ \eta_2(\bm{x},\bm{y}) = \abs{\sqrt{\hat{K}_{\hat{R}}(\bm{x})} - \sqrt{\hat{K}_{\hat{R}}(\bm{y})}}^2.$$ 
Here, $\hat{K}_{\hat{R}}(\bm{x})$ is the empirical estimator of Ripley's $K$-function \citep{ripley1976second,ripley1977modelling} defined as
$$\hat{K}_{r}(\bm{x}) = \abs{S}\sum_{\{ u, v \} \subseteq \bm{x}} \frac{\mathbbm{1}[0 < ||u - v|| \leq r]}{n(n-1)}e(u,v),$$ 
where $e(u,v)$ is an edge correction factor introduced in \citet{illian2008statistical}. 
Here we propose to leverage the ``isotropic'' edge correction that was also used by \citet{shirota2017approximate}.

The summary statistic $\mathbb{E}(\bm{\theta}|\bm{y})$ for DPPG model is an extension of the SPP case.
Here, we also follow \citet{shirota2017approximate} to use a set of $\{\hat{K}_r(\bm{x})\}$ evaluated at $10$ equally spaced values over $[0.01,0.1]$ for $\bm{\eta_2}(\bm{x},\bm{y})$, that is, 
$$\bm{\eta_2}(\bm{x},\bm{y}) = ( \eta_{2,r_1}(\bm{x},\bm{y}),\eta_{2,r_2}(\bm{x},\bm{y}), \dots, \eta_{2,r_{10}}(\bm{x},\bm{y}) )$$
over the set $(r_1 = 0.01, r_2 = 0.02,\dots, r_{10} = 0.1)$, where $\eta_{2,r_i}(\bm{x},\bm{y}) = \abs{\sqrt{\hat{K}_{r_i}(\bm{x})}-\sqrt{\hat{K}_{r_i}(\bm{y})}}^2$ is defined in the same way as the SPP case. \\

The generalized linear regression under a multi-response Gaussian family is fit with lasso regression, and cross-validation is applied to determine the penalty parameter for the lasso. 
After obtaining $\bm{\hat{a}}$ and $\bm{\hat{b}}$ using $L$ pilot draws, the distance measures $\{ \Psi(\bm{\hat{\theta}}_l, \bm{\hat{\theta}_{\text{obs}}}) \}^{L}_{l = 1}$ can be calculated in order for setting the acceptance threshold $\epsilon$ in \citet{shirota2017approximate} ABC-MCMC algorithm as shown in Algorithm~\ref{SGABC_algorithm} by taking $p$ percent estimated percentile of those measures. 
Here, the value of $p$ is set by practitioners prior to implementations.
The measure $\Psi(\bm{\hat{\theta}}_l, \bm{\hat{\theta}}_{\text{obs}}) := \sum_i (\hat{\theta}_{l,i} - \hat{\theta}_{\text{obs},i})^2/\hat{\text{var}}(\hat{\theta}_i)$ is the component-wise sum of quadratic loss for the log parameter vector with $\bm{\hat{\theta}_{\text{obs}}} = \bm{\hat{a}}$, and $\hat{\text{var}}(\hat{\theta}_i)$ is the sample variance of the $i$th component of $\bm{\hat{\theta}}$.
\begin{algorithm}[h]
\caption{\citet{shirota2017approximate} ABC-MCMC algorithm}
\label{SGABC_algorithm}
\begin{algorithmic} 
\State First initialise $\bm{\theta}^{(0)}$.
\For {$t=1$ to $T$}
\State 1. Generate $\bm{\theta}' \sim p(\cdot|\bm{\theta}^{(t-1)})$ and $\bm{x}' \sim \mathcal{L}(\cdot|\bm{\theta}')$ and calculate $\bm{\hat{\theta}}' = \bm{\hat{a}} + \bm{\hat{b}}\bm{\eta}(\bm{x}',\bm{y})$. \\
\hspace{2.4em} Repeat this step until $\Psi(\bm{\hat{\theta}}', \bm{\hat{\theta}_{\text{obs}}}) \leq \epsilon$.
\State 2. Set $\bm{\theta}^{(t)} = \bm{\theta}'$ with probability $\alpha_{\scalebox{.85}{$\scriptscriptstyle S\&G$}} = \text{min}\left(1, \frac{\pi(\bm{\theta}')p(\bm{\theta}^{(t-1)}|\bm{\theta}')}{\pi(\bm{\theta}^{(t-1)})p(\bm{\theta}'|\bm{\theta}^{(t-1)})}\right).$\\
\hspace{2.4em} Otherwise, set $\bm{\theta}^{(t)} = \bm{\theta}^{(t-1)}$.
\EndFor
\end{algorithmic}
\end{algorithm}

\subsection{Correcting the \citet{shirota2017approximate} ABC-MCMC algorithm}

Here we explain that the \citet{shirota2017approximate} ABC-MCMC Algorithm~\ref{SGABC_algorithm} does not follow exactly the ABC-MCMC scheme proposed in \citet{marjoram2003markov} or the semi-automatic ABC-MCMC algorithm proposed by \citet{fearnhead2012constructing} as shown in Algorithm~\ref{FPABCMCMC_algorithm}.

\begin{algorithm}[h]
\caption{\citet{fearnhead2012constructing} ABC-MCMC algorithm}
\label{FPABCMCMC_algorithm}
\begin{algorithmic} 
\State First initialise $\bm{\theta}^{(0)}$.
\For {$t=1$ to $T$}
\State 1. Generate $\bm{\theta}' \sim p(\cdot|\bm{\theta}^{(t-1)})$ and $\bm{x}' \sim \mathcal{L}(\cdot|\bm{\theta}')$ and calculate $\bm{\hat{\theta}}' = \bm{\hat{a}} + \bm{\hat{b}}\bm{\eta}(\bm{x}',\bm{y})$. \\
\hspace{2.4em} If $\Psi(\bm{\hat{\theta}}', \bm{\hat{\theta}_{\text{obs}}}) \leq \epsilon$, go to Step $2$. Otherwise, set $\bm{\theta}^{(t)} = \bm{\theta}^{(t-1)}$ and skip Step $2$.
\State 2. Set $\bm{\theta}^{(t)} = \bm{\theta}'$ with probability $\alpha = \text{min}\left(1, \frac{\pi(\bm{\theta}')p(\bm{\theta}^{(t-1)}|\bm{\theta}')}{\pi(\bm{\theta}^{(t-1)})p(\bm{\theta}'|\bm{\theta}^{(t-1)})}\right).$\\
\hspace{2.4em} Otherwise, set $\bm{\theta}^{(t)} = \bm{\theta}^{(t-1)}$.
\EndFor
\end{algorithmic}
\end{algorithm}

Instead of an accept-or-stay step shown as Step $1$ of Algorithm~\ref{FPABCMCMC_algorithm}, \citet{shirota2017approximate} propose to leverage an ABC-like rejection sampling step as a proposal step within the Metropolis-Hastings algorithm, that is, Step $1$ of Algorithm~\ref{SGABC_algorithm}. 
The corresponding proposal density thus is
\begin{equation}
\label{SGABCProposal}
\begin{split}
p_{\epsilon}(\bm{\theta}',\bm{x}'|\bm{\theta})
&= \frac{p(\bm{\theta}'|\bm{\theta})\mathcal{L}(\bm{x}'|\bm{\theta}')\mathbbm{1}_{A_{\epsilon,\bm{y}}}(\bm{x}')}{\int_{A_{\epsilon,\bm{y}}\cross\bm{\Theta}''}p(\bm{\theta}''|\bm{\theta})\mathcal{L}(\bm{x}''|\bm{\theta}'')\text{d}\bm{x}''\text{d}\bm{\theta}''},\\ 
\end{split}
\end{equation}
where the set $A_{\epsilon,\bm{y}}:=\left\{\bm{x}''\in\bm{N}_f: \Psi\left(\bm{\hat{a}} + \bm{\hat{b}}\bm{\eta}(\bm{x}'',\bm{y}), \bm{\hat{\theta}_{\text{obs}}}\right) \leq \epsilon\right\}$, and $\bm{\Theta}''$ denotes the parameter space over which $\bm{\theta}''$ is defined.
The indicator function $\mathbbm{1}_{A}(\bm{x}')$ gives $1$ if $\bm{x}'\in A$ and gives $0$ otherwise.
Note that the proposal density (\ref{SGABCProposal}) involves a normalizing term which we denote as 
$\zeta(\bm{\theta})=\int_{A_{\epsilon,\bm{y}}\cross\bm{\Theta}''}p(\bm{\theta}''|\bm{\theta})\mathcal{L}(\bm{x}''|\bm{\theta}'')\text{d}\bm{x}''\text{d}\bm{\theta}''$ and which is intractable in general, a point apparently missed by \citet{shirota2017approximate}.
Note also that \citet{shirota2017approximate} do not explicitly detail the ergodic distribution resulting from their algorithm. 
However, one can see from the acceptance probability in Step 2 of Algorithm~\ref{SGABC_algorithm}, that the target distribution must be written as
\begin{equation}
\label{SGABCTarget}
\begin{split}
\hat{\pi}_{\epsilon}(\bm{\theta},\bm{x}|\bm{y})
&\propto \pi(\bm{\theta})\mathcal{L}(\bm{x}|\bm{\theta})\mathbbm{1}_{A_{\epsilon,\bm{y}}}(\bm{x})\zeta(\bm{\theta}).\\ 
\end{split}
\end{equation}
This is to ensure that the ratio in $\alpha_{\scalebox{.85}{$\scriptscriptstyle S\&G$}}$ in Algorithm~\ref{SGABC_algorithm} (comprised of the usual target ratio multiplied by proposal ratio) appears as:
\begin{equation}
\label{SGABCAR}
\begin{split}
\frac{p(\bm{\theta}|\bm{\theta}')\mathcal{L}(\bm{x}|\bm{\theta})\mathbbm{1}_{A_{\epsilon,\bm{y}}}(\bm{x})/\zeta(\bm{\theta}')}{p(\bm{\theta}'|\bm{\theta})\mathcal{L}(\bm{x}'|\bm{\theta}')\mathbbm{1}_{A_{\epsilon,\bm{y}}}(\bm{x}')/\zeta(\bm{\theta})}\times\frac{\pi(\bm{\theta}')\mathcal{L}(\bm{x}'|\bm{\theta}')\mathbbm{1}_{A_{\epsilon,\bm{y}}}(\bm{x}')\zeta(\bm{\theta}')}{\pi(\bm{\theta})\mathcal{L}(\bm{x}|\bm{\theta})\mathbbm{1}_{A_{\epsilon,\bm{y}}}(\bm{x})\zeta(\bm{\theta})} 
&= \frac{\pi(\bm{\theta}')p(\bm{\theta}|\bm{\theta}')}{\pi(\bm{\theta})p(\bm{\theta}'|\bm{\theta})}
.
\end{split}
\end{equation}
As a consequence, the target distribution (\ref{SGABCTarget}) resulting from Algorithm~\ref{SGABC_algorithm} differs from the target distribution, $\pi_{\epsilon}(\bm{\theta},\bm{x}|\bm{y})\propto\pi(\bm{\theta})\mathcal{L}(\bm{x}|\bm{\theta})\mathbbm{1}_{A_{\epsilon,\bm{y}}}(\bm{x})$, resulting from Algorithm~\ref{FPABCMCMC_algorithm}, 
 by a multiplicative term, $\zeta(\bm{\theta})$, which itself depends on the model parameter(s) $\bm{\theta}$. 
Further, it is not clear what effect the term $\zeta(\bm{\theta})$ has on the target distribution.

Table~\ref{Comp_SGandFP_ABCMCMC_SPPTable} shows the comparisons of accuracy performance between the Algorithm~\ref{SGABC_algorithm} and the Algorithm~\ref{FPABCMCMC_algorithm}.
The implementations are based on our SPP simulation study settings introduced in Section~\ref{SS1SPP}, where the algorithms are implemented on a SPP for the same number of iterations.
The Ground Truth (GT) benchmark corresponds to a very long run of the exchange algorithm.
Here, 3 different values of $p=2.5, 1, 0.5$ are considered, respectively, in different implementations and for both algorithms.
It is shown that the S\&G ABC-MCMC algorithm provides worse accuracy performance for each $p$ setting compared to the F\&P ABC-MCMC algorithm, though the computational run time is much longer.
Similar deteriorated performance of the S\&G ABC-MCMC algorithm compared to the F\&P one can also be observed for the implementations based on our DPPG simulation study settings proposed in Section~\ref{SS2DPPG}.
Though the output are not detailed here, materials can be provided upon request.
Note further that the illustrated results are generally robust to multiple implementations considering the scale of the corresponding model parameters.
\begin{table}[h]
\begin{center}
\scalebox{0.875}{
\begin{tabular}{ c c c c c c c c} 
\toprule
		  		    &Time(Min)	   & E($\beta$)    & sd($\beta$)	&  $\lvert \text{Bias}(\beta)\lvert$ & E($\gamma$)	& sd($\gamma$)  & $\lvert \text{Bias}(\gamma)\lvert$ \\
\midrule
\textbf{GT} 			    & $\bm{147.55}$ & $\bm{169.13}$ & $\bm{27.669}$ &  	 	--	& $\bm{0.1339}$  & $\bm{0.0647}$  & --\\ 
\rowcolor{black!10} F\&P p2.5 &  38.482  &  171.43  &  33.394  &  2.2934  &  0.1494  &  0.0874  &  0.0155 \\
S\&G p2.5 & 138.57 & 173.26 & 28.513 & 4.1219 & 0.1297 & 0.0709 & 0.0042 \\
\rowcolor{black!10} F\&P p1 &  36.717  &  171.49  &  29.908  &  2.3565  &  0.1349  &  0.0720  &  0.0010  \\
S\&G p1 & 224.74 & 172.04 & 26.468 & 2.9080 & 0.1230 & 0.0650 & 0.0108 \\
\rowcolor{black!10} F\&P p0.5 &  35.390  &  168.53  &  28.345 &  0.6025  &  0.1396  &  0.0717  &  0.0057 \\
S\&G p0.5 & 341.93 & 170.76 & 25.836 & 1.6300 & 0.1250 & 0.0619 & 0.0089 \\
\bottomrule
\end{tabular}}
\caption{Performance comparisons between the \citet{shirota2017approximate} (S\&G) ABC-MCMC algorithm and the \citet{fearnhead2012constructing} (F\&P) ABC-MCMC algorithm. 
The implementations are based on our SPP simulation study provided in Section~\ref{SS1SPP}. 
Bold values correspond to the Ground Truth (GT), that is, a very long run of the exchange algorithm.
Here, E($\cdot$), sd($\cdot$), and $\lvert \text{Bias}(\cdot)\lvert$, respectively, represents the corresponding posterior mean, posterior standard deviation, and absolute value of bias for each model parameter.}
\label{Comp_SGandFP_ABCMCMC_SPPTable}
\end{center}
\end{table}

In order to target the correct distribution $\pi_{\epsilon}(\bm{\theta},\bm{x}|\bm{y})$, the ratio in $\alpha_{\scalebox{.85}{$\scriptscriptstyle S\&G$}}$ from Algorithm~\ref{SGABC_algorithm} should instead be modified as
\begin{equation}
\label{IntractableAlphaRatio}
\begin{split}
\frac{\pi(\bm{\theta}')\mathcal{L}(\bm{x}'|\bm{\theta}')\mathbbm{1}_{A_{\epsilon,\bm{y}}}(\bm{x}')}{\pi(\bm{\theta})\mathcal{L}(\bm{x}|\bm{\theta})\mathbbm{1}_{A_{\epsilon,\bm{y}}}(\bm{x})} \times
\frac{p(\bm{\theta}|\bm{\theta}')\mathcal{L}(\bm{x}|\bm{\theta})\mathbbm{1}_{A_{\epsilon,\bm{y}}}(\bm{x})/\zeta(\bm{\theta}')}{p(\bm{\theta}'|\bm{\theta})\mathcal{L}(\bm{x}'|\bm{\theta}')\mathbbm{1}_{A_{\epsilon,\bm{y}}}(\bm{x}')/\zeta(\bm{\theta})}
& = \frac{\pi(\bm{\theta}')p(\bm{\theta}|\bm{\theta}')\zeta(\bm{\theta})}{\pi(\bm{\theta})p(\bm{\theta}'|\bm{\theta})\zeta(\bm{\theta}')},
\end{split}
\end{equation}
which leads to what we term the corrected \citet{shirota2017approximate} ABC-MCMC algorithm shown as Algorithm~\ref{Corrected_SGABC_algorithm}.
\begin{algorithm}
\caption{Corrected \citet{shirota2017approximate} ABC-MCMC algorithm}
\label{Corrected_SGABC_algorithm}
\begin{algorithmic} \State First initialize $\bm{\theta}^{(0)}$.
\For {$t=1$ to $T$}
\State 1. Generate $\bm{\theta}' \sim p(\cdot|\bm{\theta}^{(t-1)})$ and $\bm{x}' \sim \mathcal{L}(\cdot|\bm{\theta}')$ and calculate $\bm{\hat{\theta}}' = \bm{\hat{a}} + \bm{\hat{b}}\bm{\eta}(\bm{x}',\bm{y})$. \\
\hspace{2.4em} Repeat this step until $\Psi(\bm{\hat{\theta}}', \bm{\hat{\theta}_{\text{obs}}}) \leq \epsilon$.
\State 2. Set $\bm{\theta}^{(t)} = \bm{\theta}'$ with probability $\alpha = \text{min}\left(1,  \frac{\pi(\bm{\theta}')p(\bm{\theta}^{(t-1)}|\bm{\theta}')\zeta(\bm{\theta}^{(t-1)})}{\pi(\bm{\theta}^{(t-1)})p(\bm{\theta}'|\bm{\theta}^{(t-1)})\zeta(\bm{\theta}')}\right).$\\
\hspace{2.4em} Otherwise, set $\bm{\theta}^{(t)} = \bm{\theta}^{(t-1)}$.
\EndFor
\end{algorithmic}
\end{algorithm}
However, the acceptance probability (\ref{IntractableAlphaRatio}) is generally intractable due to the intractability of the terms $\zeta(\bm{\theta})$ and $\zeta(\bm{\theta}')$.
Thus, in general, Algorithm~\ref{Corrected_SGABC_algorithm} is impossible to implement in practice.

\begin{Remark}
\normalfont  In restrictive situations, (\ref{IntractableAlphaRatio}) can become tractable.
Here, if an independent proposal distribution $p(\cdot|\bm{\theta})=p(\cdot|\bm{\theta}')=p(\cdot)$ is proposed for each iteration $t$ in Algorithm~\ref{Corrected_SGABC_algorithm}, the intractable terms $\zeta(\bm{\theta})$ and $\zeta(\bm{\theta}')$ coincide since,
$$\zeta(\bm{\theta})=\int_{A_{\epsilon,\bm{y}}\cross\bm{\Theta}''}p(\bm{\theta}'')\mathcal{L}(\bm{x}''|\bm{\theta}'')\text{d}\bm{x}''\text{d}\bm{\theta}''=\zeta(\bm{\theta}').$$
Thus both terms cancel in the corresponding acceptance ratio, forming a special case of the corrected \citet{shirota2017approximate} ABC-MCMC Algorithm~\ref{Corrected_SGABC_algorithm}.
In the case that the proposal distribution $p(\cdot)$ is set as the prior distribution $\pi(\bm{\theta})$, the resulting algorithm becomes the semi-automatic ABC algorithm introduced in \citet{fearnhead2012constructing}.
\end{Remark}

\subsection{Implementing the corrected \citet{shirota2017approximate} ABC-MCMC algorithm with Monte Carlo approximations}

Note that it is possible to approximate the intractable term $\zeta(\bm{\theta})$ in Algorithm~\ref{Corrected_SGABC_algorithm} by considering that 
\begin{equation}
\label{ApproxZeta}
\begin{split}
\zeta(\bm{\theta})=\int_{A_{\epsilon,\bm{y}}\cross\bm{\Theta}''}p(\bm{\theta}''|\bm{\theta})\mathcal{L}(\bm{x}''|\bm{\theta}')\text{d}\bm{x}''\text{d}\bm{\theta}''=\mathbbm{E}_{p(\bm{\theta}''|\bm{\theta})}[\mathbbm{E}_{\mathcal{L}(\bm{x}''|\bm{\theta}'')}(\mathbbm{1}_{A_{\epsilon,\bm{y}}}(\bm{x}''))],
\end{split}
\end{equation}
where Monte Carlo approximations can estimate the double expectation in Eq.~\eqref{ApproxZeta} based on $J_{\bm{\theta}''}$ auxiliary draws of $\bm{\theta}''\sim p(\cdot|\bm{\theta})$ and further $J_{\bm{x}''}$ auxiliary draws of $\bm{x}''|\bm{\theta}''\sim \mathcal{L}(\cdot|\bm{\theta}'')$ for each auxiliary $\bm{\theta}''$.
Though these approximations are computationally intensive, in general, we remark that these auxiliary draws can be implemented in parallel, providing some potential efficiency. 
In order to efficiently explore the performance of the corrected S\&G ABC-MCMC algorithm, we propose that the total number of the above auxiliary draws $J_{\bm{\theta}''}\times J_{\bm{x}''} \propto I$, where $I$ is the number of processors to be used for implementations in practice. 

Further, it is also possible that hundreds of $\bm{x}'$s may need to be drawn until $\Psi(\bm{\hat{\theta}}', \bm{\hat{\theta}_{\text{obs}}}) \leq \epsilon$ is satisfied in Step 1 of Algorithm~\ref{Corrected_SGABC_algorithm}.
This also brings heavy computational burden for practical implementations, and, unfortunately, current parallel computation tools are unable to be automatically applied for such \texttt{repeat} loops.
In order to improve the efficiency of ABC methods, we propose to implement an approximate parallel computation for the \texttt{repeat} loops depending on the number of processors $I$ available for parallel computation. 
More specifically, the Step $1$ of Algorithm~\ref{Corrected_SGABC_algorithm} is modified as follows: in each iteration $t$ of the algorithm,
 \begin{itemize}
 \item 1. Generate the pairs $\left\{\left(\bm{\theta}'_i, \bm{x}'_i\right):\bm{\theta}'_i \sim p(\cdot|\bm{\theta}^{(t-1)}),\;\bm{x}'_i \sim \mathcal{L}(\cdot|\bm{\theta}'_i)\right\}^I_{i=1}$ and calculate $\{\bm{\hat{\theta}}'_i:\bm{\hat{\theta}}'_i = \bm{\hat{a}} + \bm{\hat{b}}\bm{\eta}(\bm{x}'_i,\bm{y})\}^I_{i=1}$ in parallel separately on $I$ processors. 
Repeat this step until any $\Psi(\bm{\hat{\theta}}'_i, \bm{\hat{\theta}_{\text{obs}}}) \leq \epsilon$, and then let $\bm{\theta}' = \bm{\theta}'_{i'}$, where $i' = \text{min}\{i: \Psi(\bm{\hat{\theta}'}_i, \bm{\hat{\theta}_{\text{obs}}})\leq\epsilon, i = 1,2,\dots,I\}$.
 \end{itemize}
Thus, instead of a single draw in the original Step $1$ serial \texttt{repeat} loop of Algorithm \ref{Corrected_SGABC_algorithm}, the new Step 1 illustrated above propose to have $I$ parallel draws each time, significantly boosting the speed of the algorithm.
However, one key assumption of such an approximate parallel \texttt{repeat} loop is that the simulations with smaller $i$ from each round of the $I$ draws are assumed to have been simulated earlier.
By proposing the accepted draw with the smallest $i$ as the proposed state, the whole procedure is one-to-one correspondence to the original serial \texttt{repeat} loop.

\begin{Remark}
\label{Remark_cABC_MCMC_adv}
\normalfont  Note that a typical ABC algorithm, for example, the \citet{fearnhead2012constructing} semi-automatic ABC algorithm samples the proposed $\bm{\theta}'$ from the prior distribution.
This can result in inefficient posterior sampling, that is, considerably high rejection rate in the \texttt{repeat} loop if prior distribution is not appropriate or has wide support.
In comparison, one key advantage of the corrected S\&G ABC-MCMC Algorithm~\ref{Corrected_SGABC_algorithm} is that, similar to Algorithm~\ref{FPABCMCMC_algorithm}, the proposed state is sampled dependent on the current state introducing Markov property in the posterior samples.
Though the Monte Carlo approximations applied in practice for obtaining the acceptance ratio bring heavy computational burden, this can be resolved by parallel computation on as many as possible processors that the practitioners use for implementations.
\end{Remark}


\section{Simulation Studies}
\label{SS}

In this section, we explore the performances of the Noisy M-H (NMH) algorithm and the corrected S\&G ABC-MCMC algorithm (Algorithms \ref{NMH} and \ref{Corrected_SGABC_algorithm}) by comparing to the corresponding benchmarks: the Exchange (Ex) algorithm and the F\&P ABC-MCMC algorithm (Algorithms \ref{Ex} and \ref{FPABCMCMC_algorithm}), respectively. 
Comparisons between these two different classes of approaches are also provided.
We implement the algorithms on two sets of synthetic data randomly simulated from a SPP and a DPPG, respectively.
In the DPPG simulation study, the performances of the Metropolis Hastings (M-H) Algorithm~\ref{MH} as well as the approximate Noisy M-H ($\text{NMH}^{app}$) Algorithm~\ref{ApproximateNMHforDPP} are also explored.
This includes the approximate Exchange ($\text{Ex}^{app}$) algorithm which is the $K=1$ specific case of the $\text{NMH}^{app}$ algorithm.

The (approximate) Noisy M-H algorithm implementations are mainly explored for small $K: K\leq I$, where $I=7$ is the number of available processing cores in our experiments.
This is for two reasons. 
Firstly, if $K>I$, then more computation time is needed to synchronize the output of the $I$ likelihood draws, before starting another set of $I$ auxiliary chains. 
Secondly, in practice, we find that not much additional statistical efficiency results if $K>I$.
Regarding the corrected S\&G ABC-MCMC algorithm, the number of draws $J_{\bm{\theta}''}$ and $J_{\bm{X}''}$ for the double Monte Carlo approximations in Eq.~\eqref{ApproxZeta} are set to be $I$ and $7I$, respectively.  This leads to $14I^2$ auxiliary draws in each iteration to approximate the $\zeta(\cdot)$ terms.

Several summary statistics are chosen to assess the algorithm performances. 
The effective sample size (ESS) \citep{kass1998markov} due to autocorrelation is usually defined as $\text{ESS}(\theta) = T/[1+2\sum^{\infty}_{i=1}\nu_i(\theta)]$, where $T$ is the posterior sample size, $\nu_i$ is the autocorrelation at lag $i$, and the infinity sum is often truncated at lag $i$ when $\nu_i(\theta)<0.05$. 
ESS is used to check the dependence and the autocorrelation of posterior samples, that is, the mixing performance, where larger values of ESS imply better mixing. 
Due to the fact that the computational runtime of different algorithms varies considerably, we propose to monitor the ESS per second (ESS/sec), to assess the efficiency of the algorithms. 
All experiments are based on a CPU with a 1.80GHz processor and $7$ cores. 
Some basic statistics are presented to explore accuracy performances.
These include mean, standard deviation, density plot and absolute value of bias for the posterior samples of each model parameter.

\subsection{Strauss Point Process}
\label{SS1SPP}

\begin{figure}[h!]
\centering
\includegraphics[scale=0.9]{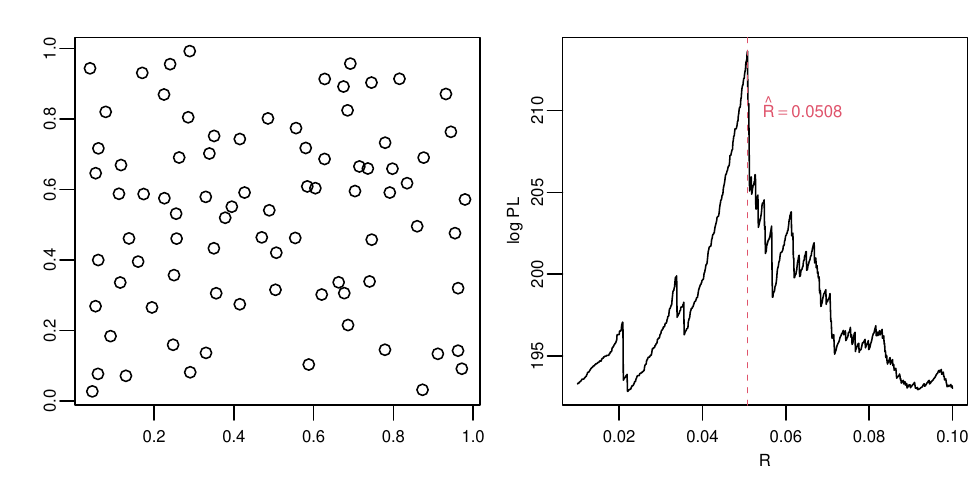}
\vspace{-10pt}
\caption{Left: Strauss point process simulated point locations $\bm{y_1}$. Right: Profile pseudo likelihood estimator $\hat{R}$ for $\bm{y_1}$.}
\label{SS1SPPobsY}
\end{figure}
Without loss of generality, we focus on the bounded spatial domain $S = [0,1]\times[0,1]$.
The SPP synthetic data $\bm{y}_1$ was randomly simulated on $S$ with parameters $\beta = 200, \gamma = 0.1$ and $R = 0.05$.
It contains 83 point locations as shown in the left plot of Figure \ref{SS1SPPobsY}. 
This simulated data size setting aims to mimic the sample size of the real dataset we focus on in Section~\ref{RDA}.
We consider three different values of $p$ for setting the acceptance threshold $\epsilon$ in both the F\&P ABC-MCMC and the corrected S\&G ABC-MCMC algorithms, namely, $2.5, 1$ and $0.5$. 
The initial states of $\beta$ and $\gamma$ parameters for all algorithms are set to be $\beta_0 = 190$ and $\gamma_0 = 0.2$. 
The prior distributions are specified to be uniform distributions: $\beta\sim\text{U}(50, 400), \gamma\sim\text{U}(0,1)$ and the interaction radius $R$ is estimated by the same profile pseudo-likelihood method as used in \citet{shirota2017approximate} with the estimated value being $\hat{R} = 0.0508$ (right plot of Figure \ref{SS1SPPobsY}). 
Bounded uniform proposals are applied for both parameters conditional on current state at each iteration $t$,
that is, the proposed state follows:
\begin{equation*}
\begin{split}
\beta' & \sim  \text{U}(\max(50,\beta^{(t-1)}-\epsilon_{\beta} ),\min(400,\beta^{(t-1)}+\epsilon_{\beta})),\\
\gamma' &\sim \text{U}(\max(0,\gamma^{(t-1)}-\epsilon_{\gamma} ),\min(1,\gamma^{(t-1)}+\epsilon_{\gamma} )),
\end{split}
\end{equation*}
with $\epsilon_{\beta}$ and $\epsilon_{\gamma}$ tuned to be $65$ and $0.16$, respectively, so that the acceptance rate of the exchange algorithm and the Noisy M-H algorithm was around $0.25$. 
The acceptance rate of the F\&P ABC-MCMC algorithm was around $0.1$, while that of the corrected S\&G ABC-MCMC Algorithm~\ref{Corrected_SGABC_algorithm} was around 0.78 compared to the very high acceptance rate 0.92 provided by the incorrect S\&G ABC-MCMC Algorithm~\ref{SGABC_algorithm}.

All the algorithms except the corrected S\&G ABC-MCMC algorithm were implemented for $0.12$ million iterations which is expected to be long enough to converge.
This is confirmed by graphical analysis of the posterior traceplots for each model parameter.
The first $0.02$ million are burn-in iterations. 
Since the corrected S\&G ABC-MCMC algorithm embeds an ABC-like reject sampling step as well as hundreds of simulations for double Monte Carlo approximations, the computational burden is significantly much heavier than all other algorithms we considered.
Thus we propose to instead implement such an algorithm for 6000 iterations with 1000 iteration-burn-in for each different $p$ setting, and we further monitor the ESS per iteration (ESS/t) without burn-in iterations to assess the quality of posterior chains.
Though the posterior samples are much smaller than those of other algorithms, the traceplots still confirm the convergence.
And it is interesting that the mixing performance (measured by ESS) of the 6000-iteration corrected S\&G ABC-MCMC $p=0.5$ algorithm is marginally better than the 120,000-iteration F\&P ABC-MCMC $p=0.5$ algorithm as shown in Table~\ref{SS1SPPTable}.

\begin{table}[h!]
\begin{center}
\scalebox{0.875}{
\begin{tabular}{ lc||ccc||ccc} 
\toprule
 				& \textbf{GT} & ${\text{F\&P}}_{\text{p2.5}}$ & ${\text{F\&P}}_{\text{p1}}$ & ${\text{F\&P}}_{\text{p0.5}}$ & ${\text{cS\&G}}_{\text{p2.5}}$ & ${\text{cS\&G}}_{\text{p1}}$ & ${\text{cS\&G}}_{\text{p0.5}}$ \\
\midrule
Time(Sec)     	 	&$\bm{8853.1}$			& 2308.9 			&    2203.0  	 	&  2123.4	 		& 12,311 & 12,571 & 12,181\\ 
\rowcolor{black!10}E($\beta$) 		&$\bm{169.13}$			&  171.43    		 &    171.49	    	 &   168.53 		& 171.43 &170.60& 169.87 \\    
sd($\beta$)     		 &$\bm{27.669}$			&   33.394   		 &     29.908    		  & 28.345 			& 30.741 &30.380& 27.944 \\  
\rowcolor{black!10}$\lvert \text{Bias}(\beta)\lvert$		&--				&  2.2934 &    2.3565 &   0.6025 & 2.2950 & 1.4691 & 0.7324 \\  
E($\gamma$)    		 &$\bm{0.1339}$			&  0.1494   		  &     0.1349  		  & 0.1396			& 0.1454 &0.1377  & 0.1382  \\    
\rowcolor{black!10}sd($\gamma$)     	 &$\bm{0.0647}$			&  0.0874 		 	&     0.0720     		  & 0.0717			& 0.0807 &0.0755 & 0.0705 \\
$\lvert \text{Bias}(\gamma)\lvert$	 &--				&  0.0155  &    0.0010  &   0.0057 & 0.0115 & 0.0038 & 0.0044  \\           
\rowcolor{black!10}ESS(Ave)     		 &$\bm{60075}$			&  2240.2 &   1796.0 & 1168.9 & 1048.5 & 997.99  & 1207.0  \\   
ESS(Ave)/s       		 &$\bm{6.7857}$			&  0.9702 &  0.8152  &  0.5504 & 0.0852 & 0.0794 & 0.0991   \\      
\rowcolor{black!10} ESS(Ave)/t       		 &$\bm{0.0601}$			&  0.0224 &   0.0180  &  0.0117 & 0.2097 & 0.1996 & 0.2414   \\      
\midrule
				& Ex 	& $\text{NMH}_{\text{K2}}$ & $\text{NMH}_{\text{K3}}$ & \multicolumn{1}{c}{$\text{NMH}_{\text{K4}}$} & $\text{NMH}_{\text{K5}}$ & $\text{NMH}_{\text{K6}}$ & $\text{NMH}_{\text{K7}}$\\
\midrule
Time(Sec)       		& 771.72   & 1228.7 & 1586.0			& \multicolumn{1}{c}{1783.2}	 & 1989.1 & 2025.1 			& 2497.9 		\\
\rowcolor{black!10}E($\beta$)    		& 168.98 & 169.07 		&  168.91	  &  \multicolumn{1}{c}{169.21} &  169.21 			&  169.36 			&  169.11 		\\ 
sd($\beta$)      		 & 27.745	 & 27.123 &  27.198			&   \multicolumn{1}{c}{27.080}			&   27.050 			&  27.084 			&   27.325 		\\
\rowcolor{black!10}$\lvert \text{Bias}(\beta)\lvert$	& 0.1513  & 0.0600 & 0.2236 &  \multicolumn{1}{c}{0.0776}    	&  0.0807  	    	 &  0.2281	&0.0283  		\\
E($\gamma$)   		& 0.1346  & 0.1335 & 0.1358			&   \multicolumn{1}{c}{0.1347}			&    0.1341			&   0.1347			&    0.1348		\\
\rowcolor{black!10}sd($\gamma$)  & 0.0662 & 0.0636 & 0.0657  &  \multicolumn{1}{c}{0.0638}			&    0.0632			&    0.0637			&    0.0641		\\
$\lvert \text{Bias}(\gamma)\lvert$	 &0.0008 & 0.0004& 0.0019 &  \multicolumn{1}{c}{0.0008}  &   0.0003 	    	 &   0.0008		   	& 0.0009 \\        
\rowcolor{black!10}ESS(Ave)    &  5933.5 &  7593.3 &  8060.5 &   \multicolumn{1}{c}{8606.3} &  9247.1 &  8962.1  &  9028.7  \\
ESS(Ave)/s     		& 7.6886  & 6.1801 & 5.0824	&  \multicolumn{1}{c}{4.8264}&    4.6488 & 4.4254 &  3.6145	\\  
\rowcolor{black!10}ESS(Ave)/t     		& 0.0593  & 0.0759 & 0.0806	&  \multicolumn{1}{c}{0.0861}&    0.0925 & 0.0896 & 0.0903	\\
\bottomrule
\end{tabular}}
\caption{Strauss point process: Posterior summary statistics for the F\&P ABC-MCMC algorithm, the corrected S\&G ABC-MCMC algorithm, the exchange algorithm and the Noisy M-H algorithm.
Bold values correspond to the ground truth; 
$\lvert \text{Bias}(\cdot)\lvert$ indicates the absolute value of the bias for each model parameter;
ESS(Ave) presents the posterior ESS averaged between two model parameters;
ESS(Ave)/s provides the corresponding average ESS per sec efficiency assessment; while ESS(Ave)/t reflects the average ESS per iteration quality assessment of posterior samples.}
\label{SS1SPPTable}
\end{center}
\end{table}

Table~\ref{SS1SPPTable} displays a table of output from different candidate algorithms.
The subscriptions beginning with ``p'' correspond to the different $p$ settings for the F\&P ABC-MCMC algorithm and the corrected S\&G ABC-MCMC algorithm.
Those beginning with ``K'' correspond to the Noisy M-H algorithm with different number of auxiliary draws.
The Markov chain obtained by the exchange algorithm is known to target the true posterior distribution and thus an extra $1.2$-million-iteration implementation of this algorithm was treated as a Ground Truth (GT) with $0.2$-million iterations as burn-in.
Note that E($\cdot$) and sd($\cdot$), respectively, represents the posterior mean and posterior standard deviation. By treating the mean of the GT posterior distribution as the true posterior mean, the statistic $\lvert \text{Bias}(\cdot)\lvert$ provides an accurate estimate of the corresponding absolute value of the bias between the true posterior mean and the mean of the posterior draws from the candidate algorithm.
The term ESS(Ave) records the average posterior ESS for the two model parameters. 
While ESS(Ave)/s provides the corresponding average posterior ESS per sec, which can be used to assess the computational efficiency of each algorithm. Finally, the term ESS(Ave)/t presents the average posterior ESS per iteration, and provide a further statistic to assess the quality of the posterior chains.

Overall, we can make the following observations, based on the results presented in Table~\ref{SS1SPPTable}.
The corrected S\&G ABC-MCMC algorithm with Monte Carlo approximations provides similar accuracy performance as the F\&P ABC-MCMC algorithm.
Both algorithms are able to provide better accuracy when $p$ is smaller, but mixing performance are shown to evolve in different ways.
As $p$ decreases, the ESS(Ave) value of the corrected S\&G ABC-MCMC algorithm is improved, while that of the F\&P ABC-MCMC algorithm deteriorates significantly.
Though the heavy computational burden of the corrected S\&G ABC-MCMC algorithm leads to significantly worse ESS(Ave)/s, this is restricted to our limited number of used processors and can be easily improved if more processors are available as we discussed in \text{Remark}~\ref{Remark_cABC_MCMC_adv}.
Note once again here that the posterior sample size of the corrected S\&G ABC-MCMC algorithm is only 1/20 of that of the F\&P ABC-MCMC algorithm, but with better mixing performance for smaller $p$, leading to significantly better ESS(Ave)/t performance.

The Noisy M-H algorithm with small $K$ setting is shown to provide similar accuracy performance as the exchange algorithm, though theoretically it is not guaranteed to target the true posterior distribution.
As $K$ increases, the mixing performance of Noisy M-H algorithm tends to be better, but longer computational run time is required leading to worse ESS(Ave)/s performance, while the accuracy is not further improved.
This suggests that the gains in mixing for larger $K$ might not be worthy compared to the loss in deteriorated computational runtime.
Thus setting $K=2$ is sufficient to attain best efficiency for the Noisy M-H algorithm.

The efficiency and the accuracy are key advantages of the Noisy M-H algorithm compared to the corrected S\&G ABC-MCMC algorithm.
However, the corrected S\&G ABC-MCMC algorithm is able to provide significantly better quality (ESS(Ave)/t) of posterior samples.
Potential improved accuracy can also be attained by setting even smaller $p$ down to zero.
Though this will bring further computational burden, the efficiency is able to be improved by leveraging more processors for parallel computation.

\begin{table}[h!]
\begin{center}
\scalebox{0.74}{
\begin{tabular}{ lc||ccc||ccc} 
\toprule
 									& \textbf{GT} & ${\text{F\&P}}_{\text{p2.5}}$ & ${\text{F\&P}}_{\text{p1}}$ & ${\text{F\&P}}_{\text{p0.5}}$ & ${\text{cS\&G}}_{\text{p2.5}}$ & ${\text{cS\&G}}_{\text{p1}}$ & ${\text{cS\&G}}_{\text{p0.5}}$ \\
\midrule
Time(Sec) 								& 7118.6{\scriptsize[1620.4]} & 2005.8{\scriptsize[249.70]} & 1953.9{\scriptsize[215.75]} & 1922.4{\scriptsize[170.30]} & 10,925{\scriptsize[1970.7]} & 10,996{\scriptsize[1983.5]} & 11,529{\scriptsize[2001.8]}\\ 
\rowcolor{black!10}$\lvert \text{Bias}(\beta)\lvert$ & -- &  2.2717{\scriptsize[1.7060]} & 2.6581{\scriptsize[2.8336]} & 2.7179{\scriptsize[1.8444]} & 1.8211{\scriptsize[1.5623]} & 2.2841{\scriptsize[2.3657]} & 2.4510{\scriptsize[1.8179]} \\  
$\lvert \text{Bias}(\gamma)\lvert$	 			& -- &  0.0282{\scriptsize[0.0180]} & 0.0243{\scriptsize[0.0187]} & 0.0184{\scriptsize[0.0164]} & 0.0258{\scriptsize[0.0165]} &0.0200{\scriptsize[0.0164]} & 0.0167{\scriptsize[0.0165]}  \\            
\rowcolor{black!10}ESS(Ave)  				& 59,159{\scriptsize[1663.3]} & 2344.6{\scriptsize[405.36]} & 1294.4{\scriptsize[242.44]} & 860.29{\scriptsize[220.46]} & 961.25{\scriptsize[147.90]} & 1150.8{\scriptsize[133.68]}  & 1302.3{\scriptsize[117.63]}  \\   
ESS(Ave)/s  							&8.7084{\scriptsize[2.0061]} & 1.1825{\scriptsize[0.2411]} & 0.6653{\scriptsize[0.1238]} & 0.4506{\scriptsize[0.1162]} & 0.0913{\scriptsize[0.0264]} & 0.1087{\scriptsize[0.0274]} & 0.1158{\scriptsize[0.0226]}   \\      
\rowcolor{black!10} ESS(Ave)/t  				&0.0592{\scriptsize[0.0017]} &  0.0234{\scriptsize[0.0041]} & 0.0129{\scriptsize[0.0024]}  &  0.0086{\scriptsize[0.0022]} & 0.1922{\scriptsize[0.0296]} & 0.2301{\scriptsize[0.0267]} & 0.2604{\scriptsize[0.0235]}   \\     
\midrule
									& Ex 	& $\text{NMH}_{\text{K2}}$ & $\text{NMH}_{\text{K3}}$ & \multicolumn{1}{c}{$\text{NMH}_{\text{K4}}$} & $\text{NMH}_{\text{K5}}$ & $\text{NMH}_{\text{K6}}$ & $\text{NMH}_{\text{K7}}$\\
\midrule
Time(Sec) 								& 726.02{\scriptsize[146.88]} & 1052.4{\scriptsize[196.53]} & 1307.3{\scriptsize[255.66]}	 & \multicolumn{1}{c}{1534.6{\scriptsize[302.31]}} & 1701.0{\scriptsize[333.22]} & 1879.5{\scriptsize[374.74]} 			& 2055.1{\scriptsize[413.07]} \\
\rowcolor{black!10}$\lvert \text{Bias}(\beta)\lvert$	& 0.2804{\scriptsize[0.1355]}  & 0.2815{\scriptsize[0.1859]} & 0.1833{\scriptsize[0.1530]}  &  \multicolumn{1}{c}{0.2639{\scriptsize[0.1714]}} &  0.2526{\scriptsize[0.1750]}   &  0.2391{\scriptsize[0.1334]}	&0.3518{\scriptsize[0.2176]} \\
$\lvert \text{Bias}(\gamma)\lvert$	 			&0.0009{\scriptsize[0.0010]} & 0.0007{\scriptsize[0.0005]}& 0.0008{\scriptsize[0.0006]} &  \multicolumn{1}{c}{0.0011{\scriptsize[0.0008]}}  &   0.0009{\scriptsize[0.0005]} 	    	 &   0.0010	{\scriptsize[0.0007]} & 0.0011{\scriptsize[0.0007]} \\      
\rowcolor{black!10}ESS(Ave)    				&  6047.4{\scriptsize[292.36]} &  7211.8{\scriptsize[282.91]} &7864.3{\scriptsize[316.49]} &   \multicolumn{1}{c}{8296.7{\scriptsize[494.14]}} &  8819.3{\scriptsize[148.41]} &  9061.4{\scriptsize[364.71]}  &  9213.9{\scriptsize[437.15]}  \\
ESS(Ave)/s     							& 8.6389{\scriptsize[1.7846]}  & 7.0663{\scriptsize[1.3391]} & 6.2358{\scriptsize[1.2911]} &  \multicolumn{1}{c}{5.5888{\scriptsize[1.0846]}}&   5.3726{\scriptsize[1.1077]} & 4.9978{\scriptsize[1.0203]} &  4.6559{\scriptsize[0.9900]}	\\  
\rowcolor{black!10}ESS(Ave)/t     				& 0.0605{\scriptsize[0.0029]}  & 0.0721{\scriptsize[0.0028]} & 0.0786{\scriptsize[0.0032]}	&  \multicolumn{1}{c}{0.0830{\scriptsize[0.0049]}}&    0.0882{\scriptsize[0.0015]} & 0.0906{\scriptsize[0.0036]} &  0.0921{\scriptsize[0.0044]}	\\
\bottomrule
\end{tabular}}
\caption{Strauss point process robustness checking: Each entry corresponds to the mean value of the corresponding posterior summary statistic for the corresponding candidate algorithm over replicated simulation studies on 10 different randomly simulated point patterns from the SPP on $S$ with $\beta=200, \gamma=0.1$ and $R=0.05$.
The values in the subscript square brackets are the corresponding standard deviation over 10 replicated studies.}
\label{SS1SPPRobustnessTable}
\end{center}
\end{table}

Table~\ref{SS1SPPRobustnessTable} extends the analysis provided in Table~\ref{SS1SPPTable} by exploring the performance of the different algorithms on $10$ simulated datasets. 
It is shown that the output are generally robust to multiple implementations and replicated simulation studies and the conclusions outlined previously hold here too. 
Though even larger simulated datasets can also be considered, the conclusions stated above are expected to remain the same.
We refer the readers to \text{Code and Data} section for more details of the code and simulation setting.

\subsection{Determinantal Point Process with a Gaussian Kernel}
\label{SS2DPPG}

\begin{figure}[h!]
\centering
\includegraphics[scale=0.48]{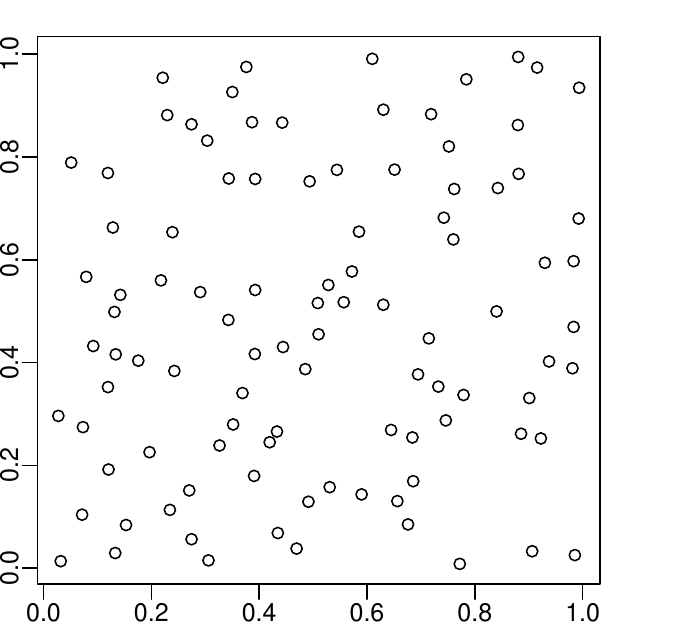}
\includegraphics[scale=0.6]{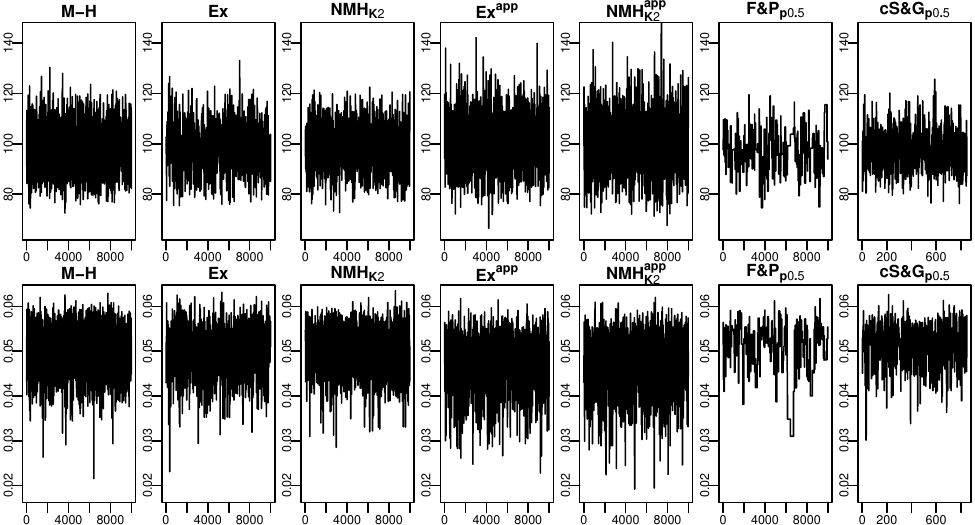}
\caption{Determinantal point process: Left: Plot of the point positions contained in the dataset $\bm{y_2}$ which was randomly generated from a DPPG. 
Right: Trace plots of the Metropolis-Hastings algorithm (M-H), the Exchange algorithm (Ex), the Noisy M-H $K=2$ algorithm ($\text{NMH}_{\text{K2}}$), the approximate Exchange algorithm ($\text{Ex}^{app}$), the approximate Noisy M-H $K=2$ algorithm ($\text{NMH}^{app}_{\text{K2}}$), the F\&P ABC-MCMC $p=0.5$ algorithm ($\text{F\&P}_{\text{p0.5}}$) and the corrected S\&G ABC-MCMC $p=0.5$ algorithm ($\text{cS\&G}_{\text{p0.5}}$).
The first and second rows correspond to the posterior samples of $\tau$ and $\sigma$, respectively. }
\label{SS2DPPG_obsYandMixing}
\end{figure}

Our second simulation study concerns a determinantal point process. 
The left plot of Figure \ref{SS2DPPG_obsYandMixing} shows the observed $n=99$ locations randomly generated from a DPP with a Gaussian kernel (DPPG) with parameters $\tau = 100, \sigma = 0.05$. 
The initial states of the two parameters in the Gaussian model are set to be $\tau_0 = 125$ and $\sigma_0 = 0.04$ and, similar to Section~\ref{SS1SPP}, uniform priors and bounded uniform proposals are proposed for the corresponding parameters, that is, $\tau\sim \text{U}(50,200)$ which includes the estimated $\hat{\tau} = n/|S|$, and $\sigma\sim \text{U}(0.001, 1/\sqrt{50\pi})$ which allows our proposed $\tau'$ to lie within our prior support. 
The proposed state in each iteration $t$ follows: $\tau' \sim \text{U}(\text{max}(50,\tau^{(t-1)}-\epsilon_{\tau} ),\text{min}(200,\tau^{(t-1)}+\epsilon_{\tau}))$ and $\sigma' \sim \text{U}(\text{max}(0.001,\sigma^{(t-1)}-\epsilon_{\sigma} ),\text{min}(1/\sqrt{\pi \tau'},\sigma^{(t-1)}+\epsilon_{\sigma} ))$ with $\epsilon_{\tau}$ and $\epsilon_{\sigma}$ tuned to be $32$ and $0.015$, respectively.

\begin{table}[h!]
\begin{center}
\scalebox{0.875}{
\begin{tabular}{ l c || c || c c || c c } 
\toprule
 				&\textbf{GT}		& M-H & ${\text{F\&P}}_{\text{p1.5}}$  & ${\text{F\&P}}_{\text{p0.5}}$ & ${\text{cS\&G}}_{\text{p1.5}}$ & ${\text{cS\&G}}_{\text{p0.5}}$\\
\midrule
Time(Sec)     	 	&\textbf{475,079}		&  19,795					& 6021.1				& 6405.3	& 110,373 & 113,143\\ 
\rowcolor{black!10}E($\tau$) 			&\textbf{98.265}	&  98.587			 			&  96.653				 	&    97.617 & 97.524 & 97.493 \\     
sd($\tau$)     		 &\textbf{7.6202}	& 7.8002						&  9.8109				&   8.5762	& 9.3970 & 8.2159\\  
\rowcolor{black!10}$\lvert \text{Bias}(\tau)\lvert$ &--	&   0.3222	 	&  1.6121					&    0.6477 & 0.7404 & 0.7713\\  
E($\sigma$)    		 &\textbf{0.0506}		& 0.0504	 				&  0.0493				&    0.0501  &    0.0503 & 0.0511 \\     
\rowcolor{black!10}sd($\sigma$)     		 &\textbf{0.0049}	& 0.0050						&  0.0071 				&   0.0066	& 0.0056 & 0.0051\\
$\lvert \text{Bias}(\sigma)\lvert$ &--	&   0.0002  	&  0.0013		&   0.0005 &  0.0003 & 0.0007 \\       
\rowcolor{black!10}ESS(Ave)    &\textbf{13,741}  &  1373.0 	 & 218.83 &    78.758 & 412.10 &  410.42\\
ESS(Ave)/s    &\textbf{0.0803} &  0.0694 	 & 0.0363 &    0.0123 &	0.0037 & 0.0036\\      
\rowcolor{black!10}ESS(Ave)/t    &\textbf{0.1374} &  0.1373 	 & 0.0219 &    0.0079 &	0.4842 & 0.4823\\   
\midrule
				& \multicolumn{1}{c}{Ex}  &  $\text{NMH}_{\text{K2}}$  & $\text{Ex}^{app}$ 		& \multicolumn{1}{c}{$\text{NMH}^{app}_{\text{K2}}$}  & $\text{NMH}^{app}_{\text{K3}}$  & $\text{NMH}^{app}_{\text{K4}}$ \\
\midrule
Time(Sec)       		& \multicolumn{1}{c}{42,456} & 46,639 & 5950.0 & \multicolumn{1}{c}{7469.3} & 8512.7  & 10,269 	  \\
\rowcolor{black!10}E($\tau$)    		&\multicolumn{1}{c}{98.524} & 98.153 	 			&  100.51			&  \multicolumn{1}{c}{100.46} 			&  100.72 			& 101.05 		 \\    
sd($\tau$)      		 &\multicolumn{1}{c}{7.8263} & 7.5343 &  9.4692 &  \multicolumn{1}{c}{10.202}  &  9.5717			 &  9.4754 		 \\
\rowcolor{black!10}$\lvert \text{Bias}(\tau)\lvert$ &\multicolumn{1}{c}{0.2590}  	& 0.1121 & 2.2432 &  \multicolumn{1}{c}{2.1931}  &   2.4508 &   2.7804		 \\
E($\sigma$)   		 &\multicolumn{1}{c}{0.0505}  & 0.0507  & 0.0480 	&   \multicolumn{1}{c}{0.0477} 	 &  0.0478 	 	&   0.0475 	  \\ 
\rowcolor{black!10}sd($\sigma$)   		  &\multicolumn{1}{c}{0.0051}  	& 0.0051	 	& 0.0056	 &  \multicolumn{1}{c}{0.0058} 	 &  0.0059 			&  0.0059 		 \\
$\lvert \text{Bias}(\sigma)\lvert$ & \multicolumn{1}{c}{0.0001} 	&  0.0015   & 0.0026 &  \multicolumn{1}{c}{0.0028} &   0.0028	 &  0.0031 \\
\rowcolor{black!10}ESS(Ave)    		 & \multicolumn{1}{c}{636.64}  & 771.70  & 836.85 &   \multicolumn{1}{c}{865.33}   &  936.92  &  1059.8  \\  
ESS(Ave)/s    		 & \multicolumn{1}{c}{0.0150}  & 0.0166   & 0.1407 &   \multicolumn{1}{c}{0.1159} 		&  0.1101  &  0.1032  \\    
\rowcolor{black!10}ESS(Ave)/t    &\multicolumn{1}{c}{0.0637} &  0.0772 	 & 0.0837 &    \multicolumn{1}{c}{0.0865}  &	0.0937 & 0.1060\\   
\bottomrule
\end{tabular}}
\caption{Determinantal point process: Posterior summary statistics for the Metropolis-Hastings algorithm, the F\&P ABC-MCMC algorithm, the corrected S\&G ABC-MCMC algorithm, the (approximate) exchange algorithm and the (approximate) Noisy M-H algorithm.
The notations with subscription being ``$app$'' correspond to the approximate exchange algorithm and the approximate Noisy M-H algorithm.
Bold values are the ground truth; $\lvert \text{Bias}(\cdot)\lvert$ indicates the absolute value of the bias for each model parameter;
ESS(Ave) presents the posterior ESS averaged between two model parameters;
ESS(Ave)/s provides the corresponding average ESS per sec efficiency assessment; while ESS(Ave)/t reflects the average ESS per iteration quality assessment of posterior samples.
}
\label{SS2DPPGTable}
\end{center}
\end{table}

In this simulation study, the approximate Noisy M-H Algorithm \ref{ApproximateNMHforDPP} ($\text{NMH}^{app}$), of which the approximate Exchange ($\text{Ex}^{app}$) algorithm is the specific $K=1$ case, is further explored.
Due to the tractability of the likelihood normalizing constant, the Metropolis-Hastings (M-H) algorithm is also available and is implemented as one of the benchmarks to compare to. 
However, considering the fact that the DPPG simulation is around 36 times slower than the SPP simulation, the computational burden required by DPPG implementations is considerably increased compared to the SPP simulation study.
Thus $12000$-iteration implementations with $2000$-iteration burn-in are instead applied for all algorithms except that the corrected S\&G ABC-MCMC algorithm was implemented for 1000 iterations with 150-iteration-burn-in.
Note that \citet{shirota2017approximate} also focuses on 1000-iteration-implementations.
Further, we propose to focus on $p=1.5$ and $0.5$ settings for both the F\&P ABC-MCMC algorithm and the corrected S\&G ABC-MCMC algorithm.
Indeed the right plot of Figure \ref{SS2DPPG_obsYandMixing} indicates that the mixing is sufficiently adequate to sample from the stationary distribution, except that the F\&P ABC-MCMC algorithm provides significantly worse mixing performance than other algorithms including the 1000-iteration corrected S\&G ABC-MCMC algorithm.
Here the M-H algorithm is additionally implemented for $0.12$ million iterations and the corresponding posterior samples with $20000$-iteration burn-in are treated as the ground truth. 

\begin{figure}[h!]
\centering
\includegraphics[scale=0.875]{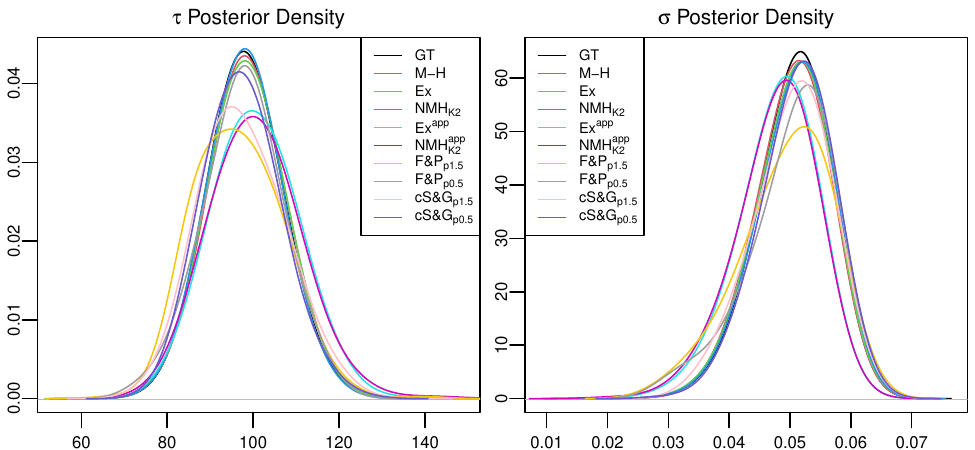}
\caption{Determinantal point process: Posterior density plots of the Ground Truth (GT), the Metropolis-Hastings algorithm (M-H), the Exchange algorithm (Ex), the Noisy M-H $K=2$ algorithm ($\text{NMH}_{\text{K2}}$), the approximate Exchange algorithm ($\text{Ex}^{app}$), the approximate Noisy M-H $K=2$ algorithm ($\text{NMH}^{app}_{\text{K2}}$), the F\&P ABC-MCMC algorithm ($\text{F\&P}_{\text{p}1.5}$, $\text{F\&P}_{\text{p}0.5}$) and the corrected S\&G ABC-MCMC algorithm ($\text{cS\&G}_{\text{p}1.5}$, $\text{cS\&G}_{\text{p}0.5}$).}
\label{SS2DPPGDensityplot}
\end{figure}

Table~\ref{SS2DPPGTable} illustrates that the comparisons between the F\&P ABC-MCMC algorithm and the corrected S\&G ABC-MCMC algorithm as well as the comparisons between the exchange algorithm and the Noisy M-H algorithm are similar to our SPP simulation study shown in Section~\ref{SS1SPP}.
The corrected S\&G ABC-MCMC algorithm can still provide the best ESS(Ave)/t performance among all the candidate algorithms.
While the efficiency (ESS(Ave)/s) of the exchange algorithm and of the Noisy M-H algorithm deteriorates a lot and is worse than that of the M-H algorithm, though these three algorithms provide similar accuracy performance.
This is due to the fact that (\lowerromannumeral{1}) the M-H algorithm is able to provide better mixing (ESS(Ave)) performance; (\lowerromannumeral{2}) the increased computational burden required by the auxiliary-draws and the likelihood function evaluation in the exchange algorithm and the Noisy M-H algorithm leads to significantly increased computational runtime. 

In comparison, our proposed approximate exchange algorithm and approximate Noisy M-H algorithm are able to provide improved computational runtime and mixing performance, leading to significantly better efficiency performance compared to the exchange algorithm and the Noisy M-H algorithm.
But the trade-off is that the accuracy is deteriorated as shown in Table \ref{SS2DPPGTable} and Figure \ref{SS2DPPGDensityplot}.
However, the accuracy performance is shown to be comparable with the F\&P ABC-MCMC $p=1.5$ algorithm and the corrected S\&G ABC-MCMC $p=1.5$ algorithm.
Thus we do not lose much accuracy when applying the approximate exchange algorithm and the approximate Noisy M-H algorithm on DPPG, while the efficiency and mixing is even better than that of the M-H algorithm.
But similar to the Noisy M-H algorithm, the biases provided by the approximate Noisy M-H algorithm is shown to not significantly reduce for larger $K$.


\section{Real Data Application}
\label{RDA}

\begin{figure}[h!]
\centering
\includegraphics[scale=0.9]{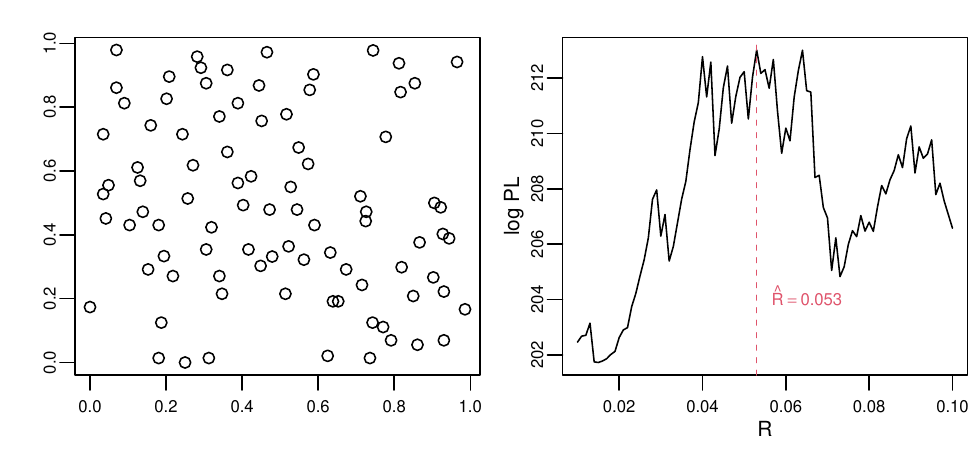}
\vspace{-10pt}
\caption{Left: Plot of the tree positions of real Duke Forest dataset $\bm{y}_{obs}$. Right: Profile pseudo likelihood estimator $\hat{R}$ for $\bm{y}_{obs}$.}
\label{RDAObsY}
\end{figure}

In this section, we apply the same experiments as our first SPP simulation study shown in Section~\ref{SS1SPP} for the purpose of algorithm exploration and comparison on a real dataset.
The original data contains $13,655$ tree locations with $68$ species in the Blackwood region of Duke Forest. 
\citet{shirota2017approximate} processed this dataset by aggregating the species and by removing trees which are under 40 dbh (diameter at breast height), considering the fact that the repulsion or inhibition can only be discovered by older trees. 
The left plot of Figure \ref{RDAObsY} illustrates the processed data which contains 89 tree locations. 
The interaction radius R within the SPP model is again estimated by the profile pseudo-likelihood method and the estimated value of $0.053$ is shown in the right plot of Figure~\ref{RDAObsY}. 
The prior settings in \citet{shirota2017approximate} are instead used here: $\beta \sim \text{U}(50, 350)$, $\gamma \sim \text{U}(0,1)$, with the bounded proposals being modified to have bounds agreeing with the prior settings. 
After the tuning process, $\epsilon_{\beta}$ and $\epsilon_{\gamma}$ are tuned to be $50$ and $0.23$, respectively. 

\begin{table}[h!]
\begin{center}
\scalebox{0.875}{
\begin{tabular}{ lc||ccc||ccc} 
\toprule
 				& \textbf{GT} & ${\text{F\&P}}_{\text{p2.5}}$ & ${\text{F\&P}}_{\text{p1}}$ & ${\text{F\&P}}_{\text{p0.5}}$ & ${\text{cS\&G}}_{\text{p2.5}}$ & ${\text{cS\&G}}_{\text{p1}}$ & ${\text{cS\&G}}_{\text{p0.5}}$ \\
\midrule
Time(Sec)     	 	&\textbf{5081.6} & 1940.4	 &    1840.2  &  1837.3	  & 9457.3 	 & 9679.2 		& 10,124  \\ 
\rowcolor{black!10}E($\beta$) 		&\textbf{143.72}		&  153.68			&  150.02			&  148.74			& 150.52 					& 151.81		& 149.48 \\     
sd($\beta$)     		 &\textbf{25.095}		&  30.551			&   28.323		&   25.697			&   30.707	 &  26.682 & 25.254 \\  
\rowcolor{black!10}$\lvert \text{Bias}(\beta)\lvert$ &--	&  9.9590 &    6.3003 &   5.0187	&6.7935 & 8.0811 & 5.7592 \\  
E($\gamma$)    		 &\textbf{0.4637}		&  0.4110			&    0.4261	        &   0.4287		&   0.4284	 &    0.4156 & 0.4202  \\      
\rowcolor{black!10}sd($\gamma$)     	 &\textbf{0.1229}	 &  0.1433	 &    0.1385		&    0.1247		&    0.1467	 &   0.1241		& 0.1201 \\
$\lvert \text{Bias}(\gamma)\lvert$ &--&  0.0527  		 &    0.0376 	    	 &   0.0350			&0.0353					& 0.0481 	 	&	0.0435  \\      
\rowcolor{black!10}ESS(Ave)  &\textbf{47,010} &  1404.0	 & 705.01	 & 441.96	  & 606.39   	& 976.71& 985.65	\\    
ESS(Ave)/s       		 &\textbf{9.2509}		&  0.7236 & 0.3831 & 0.2405	& 0.0641   & 0.1009 & 0.0974 \\   
\rowcolor{black!10} ESS(Ave)/t       		 &$\bm{0.0470}$			&  0.0140 &   0.0071  &  0.0044 & 0.1213 & 0.1953 & 0.1971   \\       
\midrule
				& Ex 	& $\text{NMH}_{\text{K2}}$ & $\text{NMH}_{\text{K3}}$ & \multicolumn{1}{c}{$\text{NMH}_{\text{K4}}$} & $\text{NMH}_{\text{K5}}$ & $\text{NMH}_{\text{K6}}$ & $\text{NMH}_{\text{K7}}$\\
\midrule
Time(Sec)       		& 449.07 & 695.24& 846.36	 & \multicolumn{1}{c}{911.71}  & 1057.4 & 1163.4  & 1370.3 	  \\
\rowcolor{black!10}E($\beta$)    		& 143.53 & 143.77& 143.92  &  \multicolumn{1}{c}{143.95}	 &  143.58	 &  143.89	 &  143.91	  \\
sd($\beta$)      		 &   24.885	 &  25.140&   24.712 &    \multicolumn{1}{c}{24.939}		&  24.947	 &   24.863	 &   24.991	  \\
\rowcolor{black!10}$\lvert \text{Bias}(\beta)\lvert$ & 0.1953 					& 0.0481 &0.1968	& \multicolumn{1}{c}{0.2279} & 0.1477	    	 &  0.1619			&0.1860 			 \\
E($\gamma$)   		&   0.4649	 &    0.4640	&   0.4645 &  \multicolumn{1}{c}{0.4642} & 0.4677 &    0.4654		&    0.4640			 \\    
\rowcolor{black!10}sd($\gamma$)   		 &    0.1234 &   0.1215&    0.1209 & \multicolumn{1}{c}{0.1221} &   0.1235			&    0.1202			&   0.1220			 \\
$\lvert \text{Bias}(\gamma)\lvert$ & 0.0012					& 0.0003 &0.0008	&  \multicolumn{1}{c}{0.0005} & 0.0040   &  0.0017		&0.0003 			 \\      
\rowcolor{black!10}ESS(Ave) & 4836.9  & 5701.5 & 6171.4		 &  \multicolumn{1}{c}{6092.5} &    6427.8&   6884.8 	&    7087.6		 \\
ESS(Ave)/s     		 & 10.771 & 8.2008 & 7.2917		 &  \multicolumn{1}{c}{6.6825} &   6.0791&   5.9177	&    5.1725	  \\   
\rowcolor{black!10}ESS(Ave)/t     		& 0.0484  &0.0570 & 0.0617	&  \multicolumn{1}{c}{0.0609}&    0.0643 & 0.0688 &  0.0709	\\
\bottomrule
\end{tabular}}
\caption{Real dataset: SPP model. Posterior summary statistics for the F\&P ABC-MCMC algorithm, the corrected S\&G ABC-MCMC algorithm, the exchange algorithm and the Noisy M-H algorithm.
Bold values correspond to the ground truth; $\lvert \text{Bias}(\cdot)\lvert$ indicates the absolute value of the bias for each model parameter;
ESS(Ave) presents the posterior ESS averaged between two model parameters;
ESS(Ave)/s provides the corresponding average ESS per sec efficiency assessment; while ESS(Ave)/t reflects the average ESS per iteration quality assessment of posterior samples.}
\label{RDATable}
\end{center}
\end{table}

\begin{figure}[b!]
\centering
\includegraphics[scale=0.87]{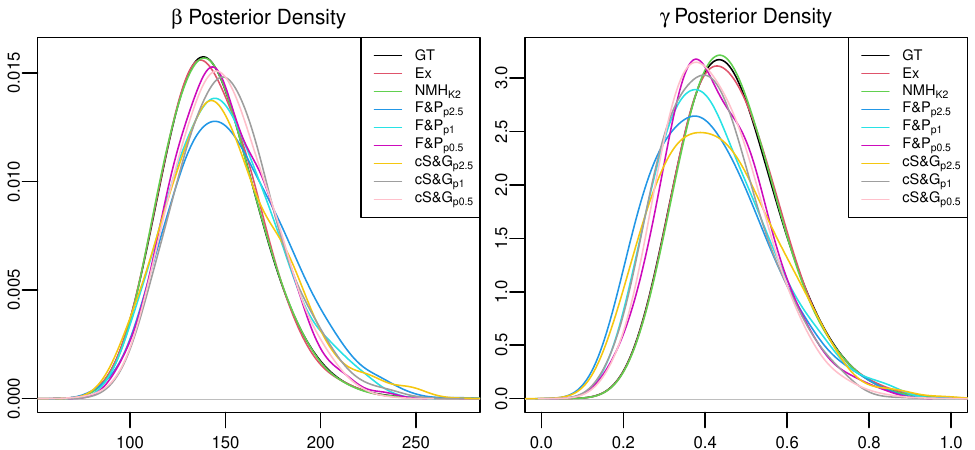}
\caption{Real dataset: Posterior density plots of the Ground Truth (GT), the Exchange algorithm (Ex), the Noisy M-H $K=2$ algorithm ($\text{NMH}_{\text{K}2}$), the F\&P ABC-MCMC algorithm ($\text{F\&P}_{\text{p}2.5}$, $\text{F\&P}_{\text{p}1}$, $\text{F\&P}_{\text{p}0.5}$) and the corrected S\&G ABC-MCMC algorithm ($\text{cS\&G}_{\text{p}2.5}$, $\text{cS\&G}_{\text{p}1}$, $\text{cS\&G}_{\text{p}0.5}$).}
\label{RDADensityplot}
\end{figure}

Our overall conclusions based on this real data experiment are broadly similar to the SPP simulation study presented in Section~\ref{SS}, where the corrected S\&G ABC-MCMC algorithm (\lowerromannumeral{1}) provides similar accuracy performance as the F\&P ABC-MCMC algorithm; (\lowerromannumeral{2}) provides better mixing compared to the F\&P ABC-MCMC algorithm for smaller $p$ setting; (\lowerromannumeral{3}) provides the best quality of posterior samples among all the algorithms including the exchange algorithm and the Noisy M-H algorithm.
However, it is shown in Table~\ref{RDATable} and Figure~\ref{RDADensityplot} that, the accuracy of both the F\&P ABC-MCMC algorithm and the corrected S\&G ABC-MCMC algorithm is not as good as that shown in our SPP simulation study, but better accuracy can be attained by proposing smaller $p$.
The Noisy M-H algorithm with small $K$ settings (\lowerromannumeral{1}) provides similar accuracy, which almost perfectly agrees with the ground truth, as the exchange algorithm, but no significantly improved accuracy observed for larger $K$; (\lowerromannumeral{2}) provides better mixing but worse efficiency for larger $K$ compared to the exchange algorithm; (\lowerromannumeral{3}) the efficiency is significantly better than that of the ABC-MCMC methods.
It is worthwhile to further note that the results we obtained for this real dataset differ from the ones provided by the incorrect ABC-MCMC algorithm illustrated in \citet{shirota2017approximate}.


\section{Conclusion and Discussion}
\label{Conclusion}

In this paper, we point out that \citet{shirota2017approximate} ABC-MCMC algorithm targets an unexpected intractable posterior distribution.
To correct such an issue, we propose a new ABC-MCMC algorithm which corrects the \citet{shirota2017approximate} ABC-MCMC algorithm, and which targets the expected stationary distribution that a typical ABC method, for example, the \citet{fearnhead2012constructing} ABC-MCMC algorithm targets.
Though the corrected \citet{shirota2017approximate} ABC-MCMC algorithm is impractical to implement generally, Monte Carlo approximations can be leveraged to estimate the intractable terms therein.
The provided exploration of such a corrected algorithm on a SPP simulation study, a DPPG simulation study and a real data application shows comparable accuracy performance and better mixing performance (for smaller thresholds) compared to the \citet{fearnhead2012constructing} ABC-MCMC algorithm.
Though the computational burden is heavy, this can be easily resolved by one using more processors for parallel computation.

The Noisy M-H algorithm with small $K$ settings we explored is shown to provide comparable accuracy performance and better mixing performance compared to the exchange algorithm.
A trade-off between the efficiency and mixing can also be observed when the number of auxiliary chains, $K$, increases.
These characteristics inherit to our proposed approximate exchange algorithm and approximate Noisy M-H algorithm, which are able to provide significantly better efficiency without much deterioration in accuracy, for DPPGs.
Note that while we do not have theoretical guarantees for these approximate algorithms, the empirical performance of them suggest that this is worthy of further study. 
In general, the (approximate) exchange algorithm and the (approximate) Noisy M-H algorithm can provide good efficiency performance, while the posterior samples provided by the corrected S\&G ABC-MCMC algorithm yield relatively very high quality of mixing.

Note that, for specific model parameter settings or sampling patterns with a large number of points, the perfect sampler can mix very slowly. However, considering that all the candidate algorithms except the M-H algorithm explored in this paper require the same perfect sampler, our overall conclusions remain the same for this situation. 
Nevertheless it would be worthwhile to further investigate the conditions under which the perfect samplers \citep[for example,][]{kendall2000perfect,lavancier2015determinantal} fail to mix efficiently in future work. Moreover, we appreciate that perfect sampling may not always be possible for more complicated point process models. This situation is beyond the scope of this article.

Future work exploring the algorithms presented in this paper on other spatial point processes can also be considered, for example, Diggle-Gratton point processes \citep{diggle1984monte}, Diggle-Gates-Stibbard processes \citep{diggle1987nonparametric}, Penttinen processes \citep{penttinen1984modelling} for which perfect simulations are known to be available.  
Note also that, in the literature, the repulsive point processes can be applied as priors to encourage separation of mixture components for mixture models, giving the repulsive mixture models \citep{beraha2023normalized,beraha2022mcmc,cremaschi2024repulsion,petralia2012repulsive,quinlan2017parsimonious}.
Since this class of models are more complex, it will be interesting to explore whether our results can be generalized to such models.

Further exploration and comparisons to other related methods are also interesting future directions.
This includes the well-known sequential ABC approaches \citep{sisson2007sequential,beaumont2009adaptive,toni2009approximate}, which can also be easily parallelized.
An ABC shadow algorithm proposed by \citet{stoica2017abc} appeared in the same year as \citet{shirota2017approximate} and was also implemented on Gibbs point processes.
Though the outputs are approximate samples from the posterior, a dependent proposal idea that is similar to the one embedded inside the corrected S\&G ABC-MCMC Algorithm~\ref{Corrected_SGABC_algorithm} was applied in their method.

\section*{Funding Statement}
The Insight Research Ireland Centre for Data Analytics is supported by Science Foundation Ireland under Grant Number 12/RC/2289$\_$P2.

\section*{Acknowledgements}

The authors would like to thank Andrew Golightly for helpful discussions. In addition, the authors gratefully thank the anonymous reviewers for their constructive comments, which helped improve this article.

\section*{Code and Data}
The application code and data for the simulation studies and the real data application is available at \url{https://github.com/Chaoyi-Lu/Bayesian_Strategies_for_RSPP}.

\bibliographystyle{apalike}
\bibliography{BayesStra_for_RSPP.bib}

\end{document}